\documentclass[useAMS,referee,usenatbib]{biom}
\usepackage{graphicx}
\usepackage{graphics}
\usepackage{amsmath}
\usepackage{amssymb}
\usepackage{pdflscape}
\newcommand\independent{\protect\mathpalette{\protect\independenT}{\perp}}
\def\independenT#1#2{\mathrel{\rlap{$#1#2$}\mkern2mu{#1#2}}}

\newcommand{\Bdist}{\mbox{Beta}}
\newcommand{\hatmu}{\widehat{\mu}}

\newcommand{\hatalpha}{\widehat{\alpha}}
\newcommand{\sumins}{\sum_{i=1}^{n_s}}
\newcommand{\as}{\alpha^*}
\newcommand{\hatphi}{\widehat{\phi}}

\newcommand{\wone}{w^{(1)}}
\newcommand{\wtwo}{w^{(2)}}

\title[Bayesian adaptive randomization in I-SPY2 SMART]{Bayesian adaptive randomization in the I-SPY2 sequential multiple assignment randomized trial}

\author{Peter Norwood$^{1,*}$\email{p.norwood@quantumleaphealth.org}, 
Christina Yau$^2$, Denise Wolf$^2$, Philip Beineke$^1$, \\ \textbf{Andrew
Chapple$^1$, Anastasios Tsiatis$^3$, and 
  Marie Davidian$^{3**}$\email{davidian@ncsu.edu}} \\
$^{1}$Quantum Leap Healthcare Collaborative, San Francisco, CA  94158,
United States\\
$^{2}$Department of Surgery, University of California, San Francisco,
San Francisco, CA 94143, United States \\
$^{3}$Department of Statistics, North Carolina State University, Raleigh, NC 27695-8203, United States\\
}

\begin{document}

% \date{{\it Received October} 2007. {\it Revised February} 2008.  {\it
% Accepted March} 2008.}

%\doi{10.1111/j.1541-0420.2005.00454.x}

%  This label and the label ``lastpage'' are used by the \pagerange
%  command above to give the page range for the article.  You may have 
%  to process the document twice to get this to match up with what you 
%  expect.  When using the referee option, this will not count the pages
%  with tables and figures.  

%\label{firstpage}

%  put the summary for your paper here - should be <= 225 words

\begin{abstract}
  The I-SPY2 phase 2 clinical trial is a long-running platform trial
  that evaluates neoadjuvant treatments for locally advanced breast
  cancer, assigning subjects to novel agents using response-adaptive
  randomization. Recently, I-SPY2 was reconfigured as a sequential
  multiple assignment randomized trial (SMART), with up to three
  stages of therapy. At the first stage, a subject is assigned to a
  tumor-subtype-specific therapy. If the subject fails to show a
  satisfactory response to the initial therapy, the subject is
  assigned to a second subtype-specific therapy, and receives a third,
  rescue therapy if response is still not achieved.  The I-SPY2 SMART
  thus evaluates entire treatment regimes that recommend therapies at
  every stage if needed.  The transition of I-SPY2 to a SMART required
  development of a response-adaptive randomization scheme that updates
  randomization probabilities at each stage, aligned with the goal of
  maximizing the number of subjects who achieve a pathological
  complete response (pCR).  We present the details of our novel
  approach, which uses Thompson sampling to update randomization
  probabilities based on the posterior probability that treatments are
  part of the optimal regime.  Empirical studies that demonstrate that
  it improves within-trial regime-specific pCR rates and recommends
  optimal regimes at rates similar to uniform, nonadaptive
  randomization.  \vspace*{0.3in}
\end{abstract}

%  Please place your key words in alphabetical order, separated
%  by semicolons, with the first letter of the first word capitalized,
%  and a period at the end of the list.
%


%  As usual, the \maketitle command creates the title and author/affiliations
%  display 

\maketitle

\section{Introduction}\label{s:intro}

The I-SPY2 phase 2 trial in patients with stage II and III high-risk
breast cancer is one of the first multicenter adaptive platform
trials %\citep{ISPY2Harrington,ISPY2Das}
\citep{ISPY2Das} and has had a major impact on trial design and the
treatment of breast cancer.  Over more than a decade, I-SPY2 has
evaluated simultaneously multiple investigational neoadjuvant agents
targeted to biologically-defined patient subtypes on the basis of the
early endpoint of pathological complete response (pCR) and identified
agents having high probability to be successful in phase 3 trials
\citep{jamapembro}.  Hallmarks of I-SPY2 are its use of
response-adaptive randomization (RAR) to
\citep{hu2006theory,berry2010bayesian,atkinson2013randomised} to
preferentially assign subjects to experimental agents demonstrating
efficacy within a subject's breast cancer subtype and its role in
establishing pCR as a strong predictor of event-free survival
\citep{ISPY2pCR} and markedly improving pCR rates in some subtypes.
In some subtypes, standard of care, while highly efficacious (with
60-70\% pCR rate), has evolved to include four or more targeted and/or
cytotoxic agents in combination, which has inspired a shift toward
development of not only agents with better efficacy but also agents
with similar efficacy and lower toxicity.

\begin{figure}
  \centering
\includegraphics[width=\textwidth]{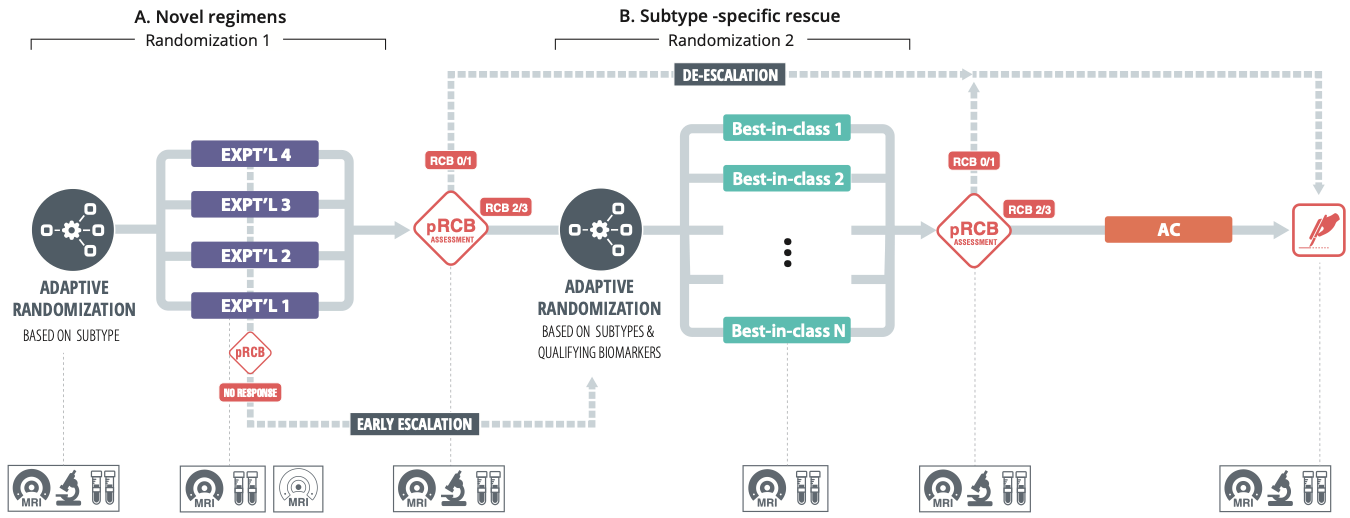}
\caption{I-SPY2 SMART schema for a given subtype.  \label{f:one}}
\end{figure}

This interest in novel targeted agents with similar efficacy to
current ``best-in-class'' agents but offering potential reduction in
toxicities inspired a reconfiguration of I-SPY2 as a sequential
multiple assignment randomized trial (SMART) \citep{murphy_2005},
referred to as I-SPY2.2 \citep{ispy_smart2,ispy_smart}, with the goal
of evaluating new agents while allowing participants deemed likely to
have a poor response to such agents the opportunity to switch to a
best-in-class agent for their tumor subtype.  Upon being classified as
belonging to a given tumor subtype, subjects follow the schema in
Figure~\ref{f:one}, involving two randomizations and three ``blocks''
of neoadjuvant therapy as follows.  Within a subtype, subjects are
initially randomized to subtype-specific novel experimental agents at
stage 1 of the SMART, referred to by the investigators as ``Block A.''
During stage 1/Block A, a subject's potential response to the assigned
agent is evaluated every 3 to 6 weeks using an imaging biomarker,
functional tumor volume (FTV), determined by breast MRI.  At the
completion of therapy (end of stage1/Block A; roughly 12 weeks),
biopsy of the tumor bed and, if needed, lymph nodes is performed, and
an algorithm referred to as ``preRCB'' incorporating FTV and biopsy
information is used to evaluate the subject's probability of pCR (pCR
can be confirmed only via surgical resection).  Subjects for whom the
probability is sufficiently high, to whom we refer as ``responders,''
are given the option to proceed to surgery without further treatment
(so meet the criteria for treatment de-escalation), at which time pCR
status is ascertained.  Those not meeting the criteria
(``nonresponders'') are recommended to continue to stage 2 of the
SMART, ``Block B,'' as are subjects who are deemed earlier in Block A
to be on a trajectory for poor response based on FTV (so eligible for
early treatment escalation), and are re-randomized to a
subtype-specific ``best-in-class'' agent.  Potential response is
evaluated during stage 2/Block B via FTV, and at the end of stage
2/Block B, following biopsy, responders meeting the preRCB
de-escalation criteria are given the option to proceed to surgery, at
which point pCR status is determined.  Nonresponders to Block B
therapy as indicated by preRCB (or poor trajectory of earlier FTV
evaluations) are recommended to receive rescue chemotherapy, at a
minimum Adriamycin (doxorubicin) and cyclophosphamide,
following which pCR status is ascertained.  For convenience, we refer
to this further therapy for stage 2/Block B nonresponders as ``stage
3,'' although it is not a true stage of the SMART, as there is
no randomization conducted.

Thus, a participant's pCR outcome is determined at the end of stage 1,
2, or 3, depending on when she proceeds to surgery.  A key feature is
that the likelihood subjects receive ineffective or unnecessary
therapy and experience potential associated toxicities is minimized.
Although we restrict attention to this current design, our proposed
methods can be generalized to a SMART with more than two
randomizations, which, in I-SPY2, would support a true stage 3 with
randomization among several rescue options (``Block C'').

% A key feature of the SMART is that that the likelihood that subjects
% receive ineffective or unnecessary therapy is minimized, so that
% exposure to potential toxicities associated with unneeded treatment is
% limited.  Thus, participants have the opportunity to avoid
% overtreatment if they are responders and to receive further,
% subtype-specific therapy if they are not.

As with any SMART, the I-SPY2 SMART design allows evaluation of entire
treatment regimes \citep{lavori_dawson,tsiatis2020dynamic}, that
dictate how therapies should be given sequentially if needed.  The
design of a SMART induces a particular set of treatment regimes,
referred to as its embedded regimes, which follow from the
randomization structure.  Here, within a specific subtype, which
determines the sets of stage 1/Block A and stage 2/Block B options, an
example of an embedded regime is ``Give experimental therapy 1; if the
patient does not proceed to surgery by the end of Block A, give
best-in-class therapy 2 followed by rescue chemotherapy if the patient
does not proceed to surgery by the end of Block B.''  Later, we denote
this regime as $\{1, 2\}$.  Evaluation of entire subtype-specific
embedded regimes is aligned with the goals of the investigators to
advance precision oncology through optimization of choice of
sequential, targeted therapies based on a patient's response in a
clinical trial design that mirrors clinical decision making in the
course of patient care.

Given the history of the successful use of Bayesian RAR in I-SPY2, a
key goal in the reconfiguration to a SMART was implementation of such
an RAR scheme to update randomization probabilities at each stage to
give greater weight to treatments that appear more efficacious based
on accrued information from previous subjects.  In conventional
randomized controlled trials (RCTs), adaptive randomization has been
argued to have the ethical benefit of improving within-trial outcomes
\citep{legocki_meurer_frederiksen_lewis_durkalski_berry_barsan_fetters_2015},
and, while there can be a loss of efficiency of post-trial inference
\citep{korn_freidlin_2011}, this concern is less critical in multi-arm
trials \citep{berry_promise}.  However, adaptive randomization is less
developed for SMARTs, which involve multiple randomizations, with only
a few approaches available
\citep{cheung_chakraborty_davidson_2014,wang_wu_wahed_2021,NorwoodRAR,YangThallWahed}.    

As the statistical team for I-SPY2, we were tasked to develop a
Bayesian RAR approach that could be introduced straightforwardly when
the trial transitioned to a two-stage SMART.  Existing methods focus
on specific designs and use ad hoc criteria for adaptation of
randomization probabilities or propose generally applicable approaches
that thus do not exploit the unique features of the I-SPY2
SMART. Accordingly, we sought to develop a fundamentally new approach
ideally suited to I-SPY2.  The approach uses Thompson sampling
\citep{thompson1933likelihood} to implement adaptation, which is a
well-known algorithm that bases randomization probabilities on the
posterior probabilities that treatment options are optimal and has
been a popular basis for RAR in conventional RCTs, particularly in
oncology \citep{berry2015brave}.  The proposed method, which is now in
use in the ongoing I-SPY2 SMART, is based on identifying for each
subtype the subtype-specific embedded regime that currently maximizes
the probability of achieving pCR.  Identifying this regime is
straightforward and leads to a computationally simple strategy for
updating approximate posterior probabilities of candidate treatments
at each stage of the SMART being optimal based on the accrued data and
thus updated randomization probabilities for each treatment option at
each stage.

In this article, we present the I-SPY2 SMART RAR strategy and
demonstrate its performance.  In Section~\ref{s:background}, we review
fundamentals of Thompson sampling, SMARTs, and treatment regimes. We
develop the statistical framework and RAR scheme in
Section~\ref{s:rand}. Section~\ref{s:sim} presents results of
simulation studies to evaluate the approach.

\section{Background}\label{s:background}

\subsection{Thompson sampling}

The central idea of Thompson sampling is that the randomization
probability for a treatment should be based on the degree of belief
that the treatment is optimal among the possible options, which is
typically represented by the posterior probability that the treatment
optimizes expected outcome
\citep{thompson1933likelihood,thall_wathen_2017}.  In I-SPY2, expected
outcome is the probability of achieving pCR. Accordingly, in a
conventional RCT, under Thompson sampling, the randomization
probability for a given treatment option $a$ in a set of possible
options $\mathcal{A}$, given the accrued data $\mathcal{D}$ to that
point, is ordinarily taken to be related to the posterior probability
$\rho(a \mid \mathcal{D})$ that $a$ optimizes expected outcome among
all options in $\mathcal{A}$.  Letting $\pi(a \mid \mathcal{D})$
denote the randomization probability based on the accrued data,
$\pi(a \mid \mathcal{D})$ is usually linked to the posterior
probability via a monotone function. A popular choice that strikes a
balance between the posterior probability itself and uniform
randomization \citep{practical_thompson} is to take
\begin{equation}
\pi(a \mid \mathcal{D})  =\frac{ \rho(a \mid \mathcal{D})^{\psi} }{ \sum_{a' \in \mathcal{A}}
  \rho(a' \mid \mathcal{D})^{\psi} },
\label{eq:randprob}
\end{equation}
where $0 \leq \psi \leq 1$ is a damping contant such that $\psi=1$
corresponds to $\pi(a \mid \mathcal{D}) = \rho(a \mid \mathcal{D})$
and $\psi=0$ yields uniform randomization. As data accumulate, the
randomization probability for an efficacious (inefficacious) treatment
will increase (decrease) toward 1 (0), so that more (fewer) subjects
will receive it. Taking $\psi$ equal to or close to 1 can result in
aggressive adaptation and randomization probabilities approaching 1 or
0 for efficacious or inefficacious treatments, respectively, which can
limit exploration of all options in $\mathcal{A}$.  Thus, an
intermediate value for $\psi$ strictly less than 1 may be a more
practical choice.  One can also take $\psi$ to be time
dependent, $\psi_t$, to allow the aggressiveness to
change with time.

In Section~\ref{s:rand}, we use the principle of Thompson sampling
based on the posterior probability that an entire embedded treatment
regime or stage 2 treatment option is optimal in the sense of
maximizing the probability of pCR to derive adaptive randomization
probabilities for subjects entering the I-SPY2 SMART at stage 1 or advancing
to stage 2.

\subsection{Treatment regimes and SMARTs}

Generically, a treatment regime is a sequence of $K$ decision rules
corresponding to $K$ decision points (stages) at which treatment
decisions are to be made, where each rule takes a subject's history of
information to that point as input and outputs a recommended treatment
option from among the available options.  As an example, consider the
regimes embedded in the I-SPY2 SMART in Section~\ref{s:intro}, for
which $K=2$.  For a given subtype $s$, $s = 1,\ldots,\mathcal{S}$,
stage 1 treatment option $a_1$ in the set of possible options
$\mathcal{A}_{1,s}$, and stage 2 treatment option $a_2$ in the set of
possible options $\mathcal{A}_{2,s}$, a subtype-specific embedded
regime has rule corresponding to stage 1 ``Give $a_1$'' and stage 2
rule ``If the patient does not proceed to surgery during/after
receiving $a_1$, give $a_2$ followed by rescue if the patient does not
proceed to surgery during/after receiving $a_2$.''  % These rules do not
% take into account additional information that may be available on an
% individual at each decision point beyond surgery status in the stage 2
% rule. 
It is possible to specify regimes with more complex rules, where the
rule at stage $k$ can incorporate baseline patient variables and
accrued covariate information up to stage $k$.  There is a vast
literature on estimating the expected outcome that would be achieved
if the patient population were to follow the rules of a regime,
referred to as the value of the regime; and on estimating an optimal
regime, one that would lead to the most beneficial expected
outcome/value in the population if the population were to follow its
rules, from suitable data.  See \citet{tsiatis2020dynamic} for a
general account.

Just as the conventional RCT is the accepted study design for
evaluating treatment effects, the SMART is the gold standard for
evaluation of treatment regimes on the basis of a given outcome.
Interest often focuses on the set of embedded regimes and on
determining the optimal embedded regime, that leading to the most
beneficial expected outcome if the patient population were to follow
its rules.  There may be more than one embedded regime achieving the
same, most beneficial expected outcome and thus more than one optimal
regime; for simplicity, we refer to ``the'' optimal embedded regime
but recognize this possibility.

In I-SPY2, the binary outcome is the indicator of achieving pCR during
the trial.  Thus, for each subtype, the optimal embedded regime is
that maximizing the overall probability of pCR.  A key concept in
evaluation of treatment regimes is that of delayed effects; namely, a
treatment option selected at an earlier stage may have implications
for which options should be selected at subsequent stages.  For
example, a Block A agent may potentiate the effect of a Block B agent
given to patients not proceeding to surgery after stage 1, so that the
Block A option maximizing the probability of pCR after stage 1 may not
be part of the regime leading to the maximum pCR rate overall.
Intuitively, adaptively updated randomization probabilities should take
possible delayed effects into account.  This objective can be achieved
by basing the randomization probability for a given option at each
stage on the probability that the option is part of an entire optimal
embedded regime.  Thus, in contrast to some approaches for two-stage
SMARTs that base the stage 1 probabilities on response status at the
end of stage 1 \citep{wang_wu_wahed_2021,YangThallWahed}, the RAR
approach we propose for I-SPY2 next is based on identifying optimal
embedded regimes for each subtype at each update of the randomization
probabilities.

\section{Bayesian adaptive randomization in the I-SPY2 SMART}\label{s:rand}

% In this section, we first describe the accrual process expected in the
% trial and how a subject's data is collected throughout that process.
% Next, we define notation for outcome (i.e. $R_k, Y_k, Y$) and accrual
% summaries like the the number of subjects have completed stage 1 from
% subtype $s$; this information is crucial for updating randomization
% probabilities. After, we cover all modeling that is required to
% generate the posterior distribution for the value of a regime. We
% conclude this section by showing how the Thompson sampling method
% updates randomization probabilities at each week.

\subsection{Statistical framework}

As is standard for conventional RCTs, we assume that subjects enter
I-SPY2 in a staggered fashion according to a completely random process
over a planned accrual period, which we take for definiteness to be 30
months.  For a given subject, upon entry, let $S$ be the subject's
subtype, and, if $S=s$, with $\mathcal{A}_{1,s}$ the set of stage 1
subtype $s$-specific novel experimental therapies, let $A_1$ be the
option from $\mathcal{A}_{1,s}$ to which the subject is randomized.
By approximately 12 weeks, whether or not a subject is a responder is
determined.  Ideally, all subjects deemed responders would proceed to
surgery; however, some responders may refuse, and nonresponders may
choose to proceed to surgery even though they are not recommended to
do so.  Accordingly, we take an intention-to-treat perspective and let
$R_1 = 1$ if the subject proceeds to surgery and 0 if not.  If
$R_1=1$, the subject's true pCR status at the end of stage 1, $Y_1$,
is ascertained, where $Y_1 = 1 (0)$ indicates that the subject did
(did not) achieve pCR, and no further data are collected on the
subject.  If $R_1 = 0$, the subject advances to stage 2 and is
randomized to a best-in-class therapy $A_2$ from among the set
$\mathcal{A}_{2,s}$ of such therapies.  By approximately 12 weeks,
whether or not the subject is a responder is determined; if the
subject proceeds to surgery, $R_2 = 1$, and $R_2=0$ otherwise.  If
$R_2=1$, the subject's true pCR status at the end of stage 2,
$Y_2 \in \{0, 1\}$, is ascertained, and no further data are collected
on the subject.  At both stages 1 and 2, there is an approximate 1
week delay between when $R_1=1$ and $R_2=1$ are observed and surgery
takes place and determination of $Y_1$ and $Y_2$ is made.  If $R_2=0$,
the subject receives rescue therapy (``stage 3''), and after
approximately 12 weeks true pCR status $Y_3 \in \{0, 1\}$ is
ascertained.  Note that $Y_k$, $k=1, 2$, is observed only if $R_k=1$,
and $Y_3$ is observed only if a subject has $R_1=R_2=0$.  The observed
data on a subject can be summarized as
$\mathcal{O} = \{ S, A_1, R_1, Y_1I(R_1=1), A_2I(R_1=0), R_2I(R_1=0),
Y_2I(R_1=0,R_2=1), Y_3I(R_1=0,R_2=0) \}$, and the primary outcome of
interest, recording whether or not pCR is achieved during the trial,
is determined by $\mathcal{O}$ as
$Y = Y_1 I(R_1=1) + Y_2 I(R_1=0, R_2=1) + Y_3 I(R_1=0, R_2=0)$, the
pCR status ascertained when surgery takes place.  Note that only one
of $Y_1, Y_2,Y_3$ is observed on a given subject.

For a subject of subtype $s$, define the following quantities.  For
$a_1 \in \mathcal{A}_{1,s}$, let
\begin{equation}
\theta_{1,s}(a_1) = P(R_1 = 1 \mid S=s, A_1=a_1), \hspace{0.1in}
\gamma_{1,s}(a_1) = P(Y_1 =1 \mid S=s, A_1=a_1,R_1=1),
\label{eq:thetagamma1}
\end{equation}
the conditional probability that a subject of subtype $s$ who receives
treatment $a_1 \in \mathcal{A}_{1,s}$ proceeds to surgery at stage
1 and the conditional probability that such a subject achieves pCR at
stage 1, respectively.  Similarly, for $a_2 \in \mathcal{A}_{2,s}$, let
\begin{equation}
\label{eq:thetagamma2}
  \begin{aligned}
&\theta_{2,s}(a_1,a_2) = P(R_2= 1 \mid S=s, A_1=a_1,R_1=0, A_2=a_2), \\
\gamma_{2,s}&(a_1,a_2) = P(Y_2 = 1 \mid S=s, A_1=a_1, R_1=0, A_2=a_2,
R_2=1),
\end{aligned}
\end{equation}
the conditional probability that a subject of subtype $s$ who received
$a_1 \in \mathcal{A}_{1,s}$, did not respond, and received treatment
$a_2 \in \mathcal{A}_{2,s}$ proceeds to
surgery at stage 2; and the conditional probability that such a subject
achieves pCR at stage 2.  Finally, the conditional probability that
a subject who did not respond at stage 2 achieves pCR at stage 3 is
\begin{equation}
\gamma_{3,s}(a_1,a_2) = P(Y_3=1 \mid S=s, A_1=a_1, R_1=0, A_2=a_2,
R_2=0).
\label{eq:gamma3}
\end{equation}

Denote by $\{a_1,a_2\}$ the subtype $s$-specific embedded regime
``Give $a_1$; if the patient does not proceed to surgery during/after
receiving $a_1$, give $a_2$ followed by rescue therapy if the patient
does not proceed to surgery during/after $a_2$'' for all possible
$(a_1,a_2) \in \mathcal{A}_{1,s} \times \mathcal{A}_{2,s}$.  Then it
can be shown that the probability of achieving pCR if the population
of subjects of subtype $s$ were to follow regime $\{a_1,a_2\}$, that
is, the value of regime $\{a_1,a_2\}$, is given by
\begin{equation}
   \label{eq:value} 
\begin{aligned}
 \mu_s(a_1,a_2) = &\theta_{1,s}(a_1) \gamma_{1,s}(a_1) +
 \{1-\theta_{1,s}(a_1)\} \theta_{2,s}(a_1,a_2) \gamma_{2,s}(a_1,a_2) \\
&+ \{1-\theta_{1,s}(a_1)\}\{1-\theta_{2,s}(a_1,a_2)\}
\gamma_{3,s}(a_1,a_2). 
\end{aligned}
\end{equation}
The expression (\ref{eq:value}) is intuitive and can be derived
formally via the g-computation algorithm; e.g., see
\citet[Section~5.4]{tsiatis2020dynamic}.  For convenience later,
denote the entire set of probabilities involved in (\ref{eq:value})
for all possible options in $\mathcal{A}_{1,s}$ and
$\mathcal{A}_{2,s}$ as
$\Theta_{1,s} = \{ \gamma_{1,s}(a_1), \gamma_{2,s}(a_1,a_2),
\gamma_{3,s}(a_1,a_2), \theta_{1,s}(a_1), \theta_{2,s}(a_1,a_2)$,
$(a_1,a_2) \in \mathcal{A}_{1,s} \times \mathcal{A}_{2,s} \}$.

From (\ref{eq:value}), for subtype $s$, the optimal embedded regime is
$\{a_{1,s}^{opt}, a_{2,s}^{opt}\}$, where $a_{1,s}^{opt}$ and
$a_{2,s}^{opt}$ jointly maximize $\mu_s(a_1,a_2)$ in $a_1$ and $a_2$.
It is straightforward to determine $a_{1,s}^{opt}$ and
$a_{2,s}^{opt}$ by evaluating (\ref{eq:value}) at all of the (finite
number of) possible combinations of options in $\mathcal{A}_{1,s}$ and
$\mathcal{A}_{2,s}$.  Note that $a_{1,s}^{opt}$ and $a_{2,s}^{opt}$
depend on $\Theta_{1,s}$.  These developments suggest that the optimal
strategy for a subject of subtype $s$ entering the SMART at any time 
during the accrual period and requiring a stage 1 treatment assignment
would be to assign the subject to  $a_{1,s}^{opt}$, the stage 1 option
associated with regime $\{ a_{1,s}^{opt}, a_{2,s}^{opt} \}$ currently
thought to be optimal and thus leading to the highest probability of
achieving pCR during the trial.  

Consider a subject of subtype $s$ already in the trial for whom
$R_1= 0$, so is eligible to receive a stage 2 treatment option, and
suppose that the subject already received a particular option $a_1$ at
stage 1, which may or may not be currently thought to be optimal in
the sense of being part of an optimal regime.  Analogous to the above,
the optimal strategy for such a subject would be to assign the subject
the option $a_{2,s}^{opt}(a_1) \in \mathcal{A}_{2,s}$, say, that maximizes
in $a_2$ the probability of achieving pCR if the population were to
receive this particular $a_1$ at stage 1 and then $a_2$ at stage 2,
given by, for fixed $a_1$,
\begin{equation}
 \mu_{2,s}(a_1,a_2) = \theta_{2,s}(a_1,a_2) \gamma_{2,s}(a_1,a_2) + \{1-\theta_{2,s}(a_1,a_2)\}
\gamma_{3,s}(a_1,a_2).
\label{eq:valuea1}
\end{equation}
It is straightforward to determine $a_{2,s}^{opt}(a_1)$ by evaluating
(\ref{eq:valuea1}) at each of the options in $\mathcal{A}_{2,s}$.  As
above, for a given $a_1$, denote the set of probabilities involved in
(\ref{eq:valuea1}), which are a subset of those in $\Theta_{1,s}$ for
the given $a_1$, as
$\Theta_{2,s}(a_1) = \{ \gamma_{2,s}(a_1,a_2),
  \gamma_{3,s}(a_1,a_2), \theta_{2,s}(a_1,a_2)$, $a_2 \in
  \mathcal{A}_{2,s} \}$.  Analogous to the above,
$a_{2,s}^{opt}(a_1)$ depends on $\Theta_{2,s}(a_1)$.

  In the next section, these observations motivate the proposed
  RAR scheme.

  \subsection{Adaptive randomization scheme}
  \label{s:adaptation}

  In principle, randomization probabilities can be updated each time a
  subject enters the trial at stage 1 or requires randomization to a
  stage 2 treatment, so at any point in continuous time.
  Logistically, however, it is practical to update the randomization
  probabilities according to a schedule; for definiteness, assume that
  randomization probabilities are updated weekly, and let $t$ index
  weeks since the beginning of enrollment.  Then for subjects who
  enter the SMART at stage 1 in the interval $[t,t+1)$ or who are
  already enrolled and reach stage 2 with $R_1=0$ during $[t,t+1)$,
  the randomization probabilities for assigning stage 1 and stage 2
  treatments updated at week $t$ would be used to assign treatments
  for these subjects.

  Denote by $\mathcal{D}_t$ the data available at week $t$ from
  subjects previously enrolled in the SMART that can be used to update
  the randomization probabilities at week $t$.  Because subjects enter
  the trial in a staggered fashion, depending how far an
  already-enrolled subject has advanced by week $t$, only a subset of
  the information in $\mathcal{O}$ will be available on the subject.

  Based on $\mathcal{D}_t$, the following summary measures can be
  calculated and used next to form relevant posterior probabilities.
  Let $n_{1,s,t}(a_1)$ denote the number of previously-enrolled
  subjects from subtype $s$ with $A_1=a_1$ who have $R_1$ observed
  before week $t$ (i.e., have completed stage 1 before week $t$).
  Among these subjects, denote by $R_{1,s,t}^{+}$ the number for whom
  $R_1=1$ before week $t$.  Let $n_{1,s,t}^{*}(a_1)$ denote the number
  of previously-enrolled subjects from subtype $s$ with $A_1=a_1$ and
  $R_1=1$ who have $Y_1$ observed before week $t$.  Finally, let
  $Y_{1,s,t}^+$ be the number of these subjects for whom $Y_1=1$
  before week $t$.

Similarly, let $n_{2,s,t}(a_1,a_2)$ be the number of previously-enrolled subjects from
subtype $s$ with $A_1=a_1, R_1=0, A_2=a_2$ who have $R_2$ observed before
week $t$ (so have completed stage 2 before week $t$), and among these let
$R_{2,s,t}^{+}$ be the number who have $R_2=1$.  Let
$n_{2,s,t}^{*}(a_1,a_2)$ denote the number of previously-enrolled subjects from subtype
$s$ with $A_1=a_1, R_1=0, A_2=a_2$, and $R_2=1$ who have $Y_2$ observed before
week $t$, and let $Y_{2,s,t}^+$ be the number of these subjects for
whom $Y_2=1$ before week $t$.

Lastly, let $n_{3,s,t}(a_1,a_2)$ be the number of previously-enrolled
subjects from subtype $s$ with $A_1=a_1, R_1=0, A_2=a_2, R_2=0$ who
have $Y_3$ observed (i.e., have completed ``stage 3'') before week
$t$, and let $Y_{3,s,t}^{+}(a_1,a_2)$ the number among these for whom
$Y_3=1$.  For convenience, a summary of the foregoing definitions is
given in  Appendix A.

We are now in a position to describe the proposed adaptive
randomization approach, which is based on the previous developments.
For a new subject of subtype $s$ entering the SMART 
during the interval $[t, t+1)$, the posterior probability that option
$\ell \in \mathcal{A}_{1,s}$ is optimal in the sense that it is
associated with the optimal subtype $s$-specific embedded regime is
\begin{equation}
\rho_{1,s,t}(\ell \mid \mathcal{D}_t) = P\{ a_{1,s}^{opt}(\Theta_{1,s})
= \ell \mid \mathcal{D}_t\},
\label{eq:rho1}
\end{equation}
where we make explicit the dependence of $a_{1,s}^{opt}$ on
$\Theta_{1,s}$.  For a subject of subtype $s$ who enrolled in the
trial during week $u < t$ and reaches stage 2 during $[t, t+1)$ and has
$R_1=0$, rather than assign the option $a_{2,s}^{opt}$ that was
determined at week $u$, given that $a_{1,s}^{opt}$ assigned at week
$u$ may or may not have been truly optimal, it makes sense to take
into account the additional data that have accrued since week $u$.
%to determine the optimal stage 2 option.
Namely, given that the subject
received option $a_1 \in \mathcal{A}_{1,s}$ at stage 1, the posterior
probability at week $t$ that option $k \in \mathcal{A}_{2,s}$ is
optimal is
\begin{equation}
\rho_{2,s,t}(k\mid a_1, \mathcal{D}_t)  = P\big[  a_{2,s}^{opt}\{a_1;
\Theta_{2,s}(a_1) \} = k \mid \mathcal{D}_t\big],
\label{eq:rho2}
  \end{equation}
where we make explicit the dependence of $a_{2,s}^{opt}(a_1)$ on
$\Theta_{2,s}(a_1)$.  

From (\ref{eq:rho1}) and (\ref{eq:rho2}), for subjects of subtype $s$,
the posterior probabilities of treatment options in
$\mathcal{A}_{1,s}$ and $\mathcal{A}_{2,s}$ being optimal at week $t$,
so based on $\mathcal{D}_t$, in the above sense thus depend on the
posterior distribution at week $t$ of $\Theta_{1,s}$, of which
$\Theta_{2,s}(a_1)$ is a subset.  To obtain this posterior
distribution at week $t$, we must specify a prior distribution for
$\Theta_{1,s}$.  A natural choice is to take the prior for each
component of $\Theta_{1,s}$ to be the $\Bdist(1, 1)$ distribution at
all $t$, which is equivalent to a uniform distribution with support
$[0,1]$.  Adoption of uniform priors for the components of
$\Theta_{1,s}$ is consistent with the principle of equipoise.
% and provides stability, so that the randomization probabilities we
% derive below do not approach 0 or 1 after only a few weeks of
% observations.  We discuss alternative choices of priors in
% Section~\ref{s:sim}

Under this specification, using the accrued data at week $t$,
$\mathcal{D}_t$, and independent uniform priors for each component of
$\Theta_{1,s}$, in obvious notation, the updated posterior
distributions at week $t$ for subjects of subtype $s$ can be obtained
as
$$\theta_{1,s}(a_1) \mid \mathcal{D}_t \sim \Bdist\left\{ 1 + R^{+}_{1,s,t}(a_1), 1 + n_{1,s,t}(a_1) - R^{+}_{1,s,t}(a_1) \right\}$$
$$\gamma_{1,s}(a_1) \mid \mathcal{D}_t \sim \Bdist\left\{1 +
  Y^{+}_{1,s,t}(a_1), 1 + n^{*}_{1,s,t}(a_1) -  Y^{+}_{1,s,t}(a_1)
\right\}$$
\begin{equation}
\theta_{2,s}(a_1,a_2) \mid \mathcal{D}_t \sim \Bdist\left\{1 +
  R^{+}_{2,s,t}(a_1,a_2), 1 + n_{2,s,t}(a_1,a_2) -
  R^{+}_{2,s,t}(a_1,a_2) \right\}
\label{eq:posteriorst}
\end{equation}
$$\gamma_{2,s}(a_1,a_2) \mid \mathcal{D}_t \sim \Bdist\left\{1 + Y^{+}_{2,s,t}(a_1,a_2), 1 + n^{*}_{2,s,t}(a_1,a_2) -  Y^{+}_{2,s,t}(a_1,a_2) \right\}$$
$$\gamma_{3,s}(a_1,a_2) \mid \mathcal{D}_t \sim \Bdist\left\{1 + Y^{+}_{3,s,t}(a_1,a_2), 1 + n^{*}_{3,s,t}(a_1,a_2) -  Y^{+}_{3,s,t}(a_1,a_2) \right\}.$$
\noindent

Given the posterior distributions (\ref{eq:posteriorst}) and the
fact that it is straightforward to draw random samples from the Beta
distribution, we propose to approximate the posterior distributions
(\ref{eq:rho1}) and (\ref{eq:rho2}) at week $t$ for subtype $s$ by
drawing a sample of size $M$, where $M$ is large, from each posterior
in (\ref{eq:posteriorst}) to obtain $M$ random draws from the joint
posterior of $\Theta_{1,s}$.  Denoting the elements of the sample as
$\Theta_{1,s}^{(m)}$, $m = 1,\ldots,M$, approximate the posterior
distribution $\rho_{1,s,t}(\ell \mid \mathcal{D}_t)$ in
(\ref{eq:rho1}) that $\ell \in \mathcal{A}_{1,s}$ is optimal by
\begin{equation}
\widehat{\rho}_{1,s,t}(\ell \mid \mathcal{D}_t) = M^{-1}\sum_{m=1}^M I\left\{
  a_{1,s}^{opt}(\Theta_{1,s}^{(m)}) = \ell \right\},
\label{eq:rho1approx}
\end{equation}
where $I(\,\cdot\,)$ is the indicator function.  Note that obtaining
(\ref{eq:rho1approx}) entails obtaining
$a_{1,s}^{opt}(\Theta_{1,s}^{(m)})$ and
$a_{2,s}^{opt}(\Theta_{1,s}^{(m)})$ jointly maximizing
(\ref{eq:value}) in $a_1$ and $a_2$ with the components of
$\Theta_{1,s}^{(m)}$ substituted for each $m = 1,\ldots,M$.
Given the approximate posterior probabilities in (\ref{eq:rho1approx})
at week $t$ for each $\ell \in
\mathcal{A}_{1,s}$, the
randomization probabilities to be used during $[t, t+1)$ for each $a_1
\in \mathcal{A}_{1,s}$ to assign stage 1 treatments to subjects of
subtype $s$ entering the SMART during $[t, t+1)$ can be calculated,
analogous to (\ref{eq:randprob}), as
\begin{equation}
\pi_{1,s,t}(a_1\mid \mathcal{D}_t; \psi_t) = \frac{\widehat{\rho}_{1,s,t}(a_1 \mid \mathcal{D}_t) ^{\psi_t}}
{ \sum_{\ell \in \mathcal{A}_{1,s}} \widehat{\rho}_{1,s,t}(\ell \mid  \mathcal{D}_t) ^{\psi_t} },
\label{eq:randprobt1}
\end{equation}
where we allow the damping constant $\psi_t$ to be week dependent.

Similarly, for each fixed $a_1 \in \mathcal{A}_{1,s}$, the posterior
distribution $\rho_{2,s,t}(k \mid \mathcal{D}_t)$ in
(\ref{eq:rho2}) that $k \in \mathcal{A}_{2,s}$ is optimal can be approximated
by
\begin{equation}
\widehat{\rho}_{2,s,t}(k \mid a_1, \mathcal{D}_t) = M^{-1}\sum_{m=1}^M I\left[
a_{2,s}^{opt}\{a_1,\Theta_{2,s}^{(m)}(a_1)\} = k \right],
\label{eq:rho2approx}
\end{equation}
where $\Theta_{2,s}^{(m)}(a_1)$ is the relevant subset of
$\Theta_{1,s}^{(m)}$.  As for (\ref{eq:rho1approx}), obtaining
(\ref{eq:rho2approx}) requires obtaining
$a_{2,s}^{opt}\{a_1,\Theta_{2,s}^{(m)}(a_1)\}$ by maximizing
(\ref{eq:valuea1}) in $a_2$ with $a_1$ held fixed and the components
of $\Theta_{2,s}^{(m)}(a_1)$ substituted for $m=1,\ldots,M$.  Given
the approximate posterior probabilities in (\ref{eq:rho2approx}) at
week $t$ for each $k \in \mathcal{A}_{2,s}$, the randomization
probabilities to be used during $[t, t+1)$ for each
$a_2 \in \mathcal{A}_{2,s}$ to assign stage 2 treatments to subjects
already enrolled in the SMART of subtype $s$ who received treatment
$a_1$ in stage 1 and are nonresponders to $a_1$ (do not proceed to
surgery) during $[t, t+1)$ can be obtained for each
$a_1 \in \mathcal{A}_{1,s}$ as
\begin{equation} 
\pi_{2,s,t}(a_2\mid a_1, \mathcal{D}_t ;\psi_t) =
\frac{\widehat{\rho}_{2,s,t}(a_2 \mid a_1, \mathcal{D}_t) ^{\psi_t}}
{\sum_{k \in\mathcal{A}_{2,s}} \widehat{\rho}_{2,s,t}(k \mid a_1, \mathcal{D}_t)^{\psi_t}}.
\label{eq:randprobt2}
\end{equation}
% The randomization probabilities in (\ref{eq:randprobt1}) and
% (\ref{eq:randprobt2}) are used to assign stage 1 treatments to
% subjects entering the SMART in the time interval $[t,t+1)$ and to
% assign stage 2 treatments to subjects reaching stage 2 during
% $[t,t+1)$ with $R_1=0$, respectively.

\citet{NorwoodRAR} propose different RAR methods for SMARTs that use
an implementation of Thompson sampling based on a frequentist analog
to a posterior distribution.  Our Bayesian approach based on
(\ref{eq:posteriorst}) is ideally suited to the I-SPY2 SMART involving
multiple subtypes and three stages of therapy.  E.g., if few patients
reach stage 3 within a subtype, fitting frequentist logistic models
for $\gamma_{3,s}(a_1,a_2)$ could be problematic, a concern that is
circumvented by sampling from the corresponding posterior in
(\ref{eq:posteriorst}).

Early in the trial, the posterior distributions (\ref{eq:posteriorst})
will not differ from the priors until some subjects have progressed
through all stages, so that the updated randomization probabilities
will differ little from those under fixed, equal randomization.  Once
adaptation commences, a concern is that an unusual configuration of
early data could lead to the randomization probabilities becoming
``stuck'' toward favoring suboptimal treatment options
\citep{thall_caveats}.  Thus, as common with RAR, a brief ``burn-in''
period of fixed, equal randomization is prudent so that sufficient
subjects can progress through all stages before adaptation is
initiated.  In the I-SPY2 SMART, such a burn-in period is used as a
precaution.

\subsection{Post-trial inference } \label{eq:postrial}

At the conclusion of the trial, when all subjects have had pCR status
ascertained, it is of interest to estimate the value of each subtype
$s$-specific embedded regime $\{a_1,a_2\}$, i.e., the pCR rate that
would be achieved if all patients in the subtype $s$ population were
to follow that regime, based on the final data $\mathcal{D}_{final}$,
say, from all subjects.  For each regime, a natural approach is to
draw a sample of size $M$ from each of the posterior distributions in
(\ref{eq:posteriorst}) to obtain draws $\Theta_{1,s}^{(m)}$,
$m=1,\ldots,M$, from the joint posterior of $\Theta_{1,s}$ given
$\mathcal{D}_{final}$ and substitute these in the expression
(\ref{eq:value}) for the value of regime $\{a_1, a_2\}$ to obtain
$\mu^{(m)}_s(a_1,a_2)$, $m = 1,\ldots,M$, which can be viewed as a
sample from the posterior distribution of $\mu_s(a_1, a_2)$.  The
Bayesian estimator for $\mu_s(a_1, a_2)$, $\hatmu_{s,Bayes}(a_1,a_2)$,
say, can then be obtained as the mode or mean of the sample, with the
standard deviation of the sample as a measure of uncertainty.  In the
simulations of Section~\ref{s:sim}, because in our experience the
distribution of the $M$ samples is approximately symmetric and
unimodal, we focus on the Bayesian estimator given by the mean of the
$M$ posterior draws,
$\hatmu_{s,Bayes}(a_1,a_2) =M^{-1} \sum^M_{m=1} \mu^{(m)}_s(a_1,a_2)$.

Frequentist-type estimators are also possible.  An intuitive plug-in
estimator for $\mu_s(a_1, a_2)$ can be obtained by estimating the
quantities in (\ref{eq:thetagamma1})-(\ref{eq:gamma3}) by functions of
the obvious sample proportions in $\mathcal{D}_{final}$, with
approximate sampling distribution obtained via standard asymptotic
normal theory; denote this estimator by $\hatmu_{s,samp}(a_1,a_2)$.
The theory is based on the assumption that the data in
$\mathcal{D}_{final}$ are independent and identically distributed
(i.i.d.) across subjects.  However, under adaptive randomization,
these data are not i.i.d. because the randomization probabilities are
functions of previous data.  Accordingly, the properties of
$\hatmu_{s,samp}(a_1,a_2)$ may not be well approximated by standard
asymptotic theory, and inference on $\mu_s(a_1,a_2)$ could be flawed
\citep{ZhangMest2021}.  Following \citet{ZhangMest2021}, one can
construct an alternative estimator we denote as
$\hatmu_{s,wtsamp}(a_1,a_2)$ in which weighted sample proportions are
used to estimate the quantities in
(\ref{eq:thetagamma1})-(\ref{eq:gamma3}), where the weights are chosen
so that asymptotic normality of $\hatmu_{s,wtsamp}(a_1,a_2)$ can be
established using the martingale central limit theorem, taking into
account the dependence in $\mathcal{D}_{final}$ induced by RAR.
Formulations of $\hatmu_{s,samp}(a_1,a_2)$ and
$\hatmu_{s,wtsamp}(a_1,a_2)$ are given in  Appendix B.

A key goal is to identify the optimal embedded regime for each subtype
$s$, denoted as before as $\{a^{opt}_{1,s}, a^{opt}_{2,s}\}$, with
value $\mu_s(a^{opt}_{1,s}, a^{opt}_{2,s})$.  The obvious estimator is
the regime with the largest estimated value (pCR rate) based on the
chosen estimator for $\mu_s(a_1, a_2)$, with
$\mu_s(a^{opt}_{1,s}, a^{opt}_{2,s})$ estimated by the corresponding
estimated pCR rate.

\section{Simulation Studies} \label{s:sim}

We carried out a suite of simulation studies under a range of
scenarios based on the I-SPY2 investigators' expectations for
experimental stage 1 agents and past data on best-in-class stage 2
agents and rescue therapy from I-SPY2.  For each scenario, we evaluate
both in-trial performance and the quality of post-trial inference for
SMARTs conducted using simple, uniform randomization at stages 1 and
2, denoted as SR; and using several versions of the proposed RAR
strategy based on Thompson sampling with different damping constants
$\psi_t$, denoted as TS($\psi_t)$, where $\psi_t$ is both time
independent and time dependent as discussed below.  To maintain the
positivity assumption under RAR and ensure adequate exploration of all
treatment options, we imposed clipping constants, i.e., lower and
upper bounds, of 0.05 and 0.95 on the adaptive probabilities
\citep{NorwoodRAR,zhang2021inference}.

All generative scenarios involve a subtype $s$ for which there are two
stage 1/Block A experimental treatment options,
$\mathcal{A}_{1,s} = \{0, 1\}$, and three stage 2/Block B
best-in-class options, $\mathcal{A}_{2,s} = \{0, 1, 2\}$.  For all
scenarios, aligned with the investigators' beliefs, achievement of pCR
was assumed to be durable over the maximum duration of a subject's
participation in the trial.  That is, if a subject achieves pCR after
stage 1, $Y_1 = 1$, but $R_1 = 0$, so that the subject proceeds to
stage 2, then $Y_2=1$.  Similarly, if a subject achieves pCR after
stage 1 or 2, so that $Y_1=1, Y_2=1$ or $Y_1=0, Y_2=1$, but
$R_1=R_2=0$, so that the subject proceeds to stage 3, then $Y_3=1$.
Based on the investigators' extensive evaluation, the
sensitivity and specificity of the preRCB algorithm were taken to be
independent of treatment and prior pCR status, as characterized
formally in  Appendix D, with sensitivity $\lambda_{sens} = 0.53$
and specificity $\lambda_{spec} = 0.90$.
% $\lambda_{sens} = P(R_1=1 \mid A_1=a_1, Y_1=1) = P(R_2=1 \mid
% A_1=a_1,Y_1=0, R_1=0, A_2=a_2,Y_2 =1) = P(R_2=1 \mid A_1=a_1,Y_1=1,
% R_1=0, A_2=a_2,Y_2 =1) = 0.53$ and
% $\lambda_{spec} = P(R_1=0 \mid A_1=a_1, Y_1=0) = P(R_2 = 0 \mid
% A_1=a_1, Y_1=0, R_1=0, A_2=a_2, Y_2=0) = P(R_2 = 0 \mid A_1=a_1,
% Y_1=1, R_1=0, A_2=a_2, Y_2=0) = 0.90$ for all
% $(a_1, a_2) \in \mathcal{A}_{1,s} \times \mathcal{A}_{2,s}$,
%

it was assumed for simplicity that all subjects would follow the
preRCB recommendation.  A given scenario involves specification of the
true pCR rates $p_1(a_1) =P(Y_1 = 1 \mid A_1 = a_1)$, $a_1 = 0, 1$,
following stage 1 treatment, and, given that pCR has not yet been
achieved following stage 1 or stage 1 and 2 treatment,
$p_2(a_1,a_2) = P(Y_2 = 1 \mid A_1 = a_1, Y_1=0, A_2=a_2)$ and
$p_3(a_1,a_2) = P(Y_3 = 1 \mid A_1=a_1, Y_1=0, A_2=a_2 Y_2=0)$,
$(a_1,a_2) \in \mathcal{A}_{1,s} \times\mathcal{A}_{2,s}$.  Under these
conditions, it is shown in  Appendix C that
the true value of regime $\{a_1, a_2\}$ is
  \begin{equation}
    \label{eq:simvalue}
    \begin{aligned}
\mu_s(a_1, a_2) = &p_1(a_1) + 
p_2(a_1,a_2)\{1-p_1(a_1)\} \lambda_{spec} \\ &+
p_3(a_1,a_2) \{1- p_2(a_1,a_2)\}\{1-p_1(a_1)\} \lambda_{spec}^2,
\end{aligned}
\end{equation}
which depends on the specificity but not on the sensitivity of the
preRCB assessment.  

Under these specifications, the simulation study for each scenario
involved 5000 Monte Carlo trials conducted under each randomization
scheme.  For each trial, $n_s$ subjects of subtype $s$ were enrolled
during an enrollment period of $T_{enroll}=130$ weeks (2.5 years),
where, for simplicity, the enrollment dates for the $n_s$ subjects
were sampled uniformly from the integers in $[1,T_{enroll}]$.  For
trials using SR, all subjects were randomized to the options in
$\mathcal{A}_{1,s}$ and $\mathcal{A}_{2,s}$ using equal probabilities
of $1/2$ and $1/3$, respectively.  For trials using RAR, a burn-in
period was implemented such that subjects enrolling through week
$t_{burn}$, where $t_{burn}$ is the week at which the 20th subject
enrolled, were randomized using SR at stages 1 and 2.  Randomization
probabilities were then updated at each week
$t = t_{burn}+1, \ldots, T_{end}$, where $T_{end}$ is the last week at
which randomization was required, and used to randomize subjects
enrolling during $[t, t+1)$ for $t \geq t_{burn}+1$.  RAR was
implemented with $\psi_t = 0.25, 0.50, 0.75, 1.00, 0.50(t/T_{end})$,
and $(t/T_{end})$; the last two choices of $\psi_t$ allow adaptation
to become more aggressive over time \citep{practical_thompson}.  For
all adaptive randomization schemes, we took $M=1000$.  Details of data
generation aligned with these conditions are presented in  Appendix D.

% Under all randomization schemes, for each subject, at enrollment,
% $A_1 \in \mathcal{A}_{1,s}$ was generated as Bernoulli using the
% current stage 1 randomization probabilities.  At week 12 post
% enrollment, $Y_1$ was generated as Bernoulli$\{p_1(a_1)\}$ for
% $A_1 = a_1$, and $R_1$ was generated as Bernoulli$\{p_R(Y_1)\}$, where
% $p_R(y) = \lambda_{sens} I(y=1) + (1-\lambda_{spec}) I(y=0)$.  If
% $R_1=1$, $Y_1$ was recorded at week 13 post enrollment, and no further
% data were generated for the subject.  If $R_1=0$,
% $A_2 \in \mathcal{A}_{2,s}$ was generated as trinomial using the
% current stage 2 randomization probabilities, and at week 25 post
% enrollment, $Y_2$ was either generated as Bernoulli$\{p_2(a_1, a_2)\}$
% for $A_1 = a_1, A_2=a_2$ if $Y_1=0$ or set equal to $Y_1$ if $Y_1=1$,
% and $R_2$ was generated as Bernoulli$\{p_R(Y_2)\}$.  If $R_2=1$, $Y_2$
% was recorded at week 26 post enrollment, and no further data were
% generated for the subject.  If $R_2=0$, at week 38 post enrollment,
% $Y_3$ was either generated as Bernoulli$\{p_3(a_1, a_2)\}$ for
% $A_1 = a_1, A_2=a_2$ if $Y_2=0$ or set equal to $Y_2$ if $Y_2=1$.
% Thus, $T_{end} = 143$ weeks.

Table~\ref{t:scenarios} summarizes the stage 1 and 2 pCR rates
used in each generative scenario; for simplicity, in all
scenarios, $p_3(a_1, a_2) = 0.15$ for all
$(a_1, a_2) \in \mathcal{A}_{1,s} \times \mathcal{A}_{2,s}$, so that
the efficacy of rescue therapy in achieving pCR at the end of stage 3
is the same regardless of the stage 1 and 2 treatments received.  In
Scenario 1, there are no delayed effects of stage 1 treatments; the
probability of achieving pCR at the end of stage 2, $Y_2 = 1$, given
$Y_1=0$, for any stage 2 treatment is the same regardless of
stage 1 treatment received.  Thus, the treatment options that maximize
the probability of pCR at the end of each of stages 1 and 2 comprise
the optimal regime $\{ 1, 0\}$.  Scenario 2 involves purely delayed
effects of stage 1 options; the probability of achieving pCR at the
end of stage 1 is the same for both options, but the
probability of pCR following stage 2 is different depending on stage 1
treatment for stage 2 options 1 and 2, so that $\{ 1, 1\}$ is the
optimal regime.  Scenarios 3 and 4 also involve delayed effects,
albeit in more complicated ways.  Scenario 3 represents an
antagonistic situation; the stage 1 option that maximizes the
probability of pCR at the end of stage 1 is associated with uniformly
lower pCR rates after any stage 2 treatment and is not associated with
the optimal regime $\{ 0, 0 \}$.  Scenario 4 involves synergy; the
stage 1 option with the lower probability of achieving pCR at the end
of stage 1 potentiates stage 2 option 0 to have a much higher pCR
rate, leading to the optimal regime $\{ 0, 0 \}$.  Scenario 5 is similar
to Scenario 3 except that both regimes $\{ 0, 0 \}$ and $\{ 0, 1 \}$
are optimal.

\begin{table}[h]
    \caption{True pCR probabilities, stages 1 and 2, and true values of
      regimes, Scenarios 1 - 5 (optimal regime in boldface).}
    \label{t:scenarios}
          \centering
  \begin{tabular}{cccccccc} \Hline
   & & &  \multicolumn{5}{c}{Scenario} \\
   \hline
  Probability of pCR & $a_1$ & $a_2$ & 1 & 2 & 3 & 4 & 5\\
  \hline
$p_1(a_1)$ & 0 &  -- &      0.30 & 0.30 & 0.30 &  0.30 &  0.30\\
                  & 1 & --  &      0.40 & 0.30 & 0.40 &  0.40 &  0.40\\
  \hline
$p_2(a_1,a_2)$ & 0 & 0 &   0.40 & 0.40 & 0.60 &  0.60 & 0.50 \\
                        & 0 & 1 &   0.30 & 0.30 & 0.50 &  0.25 & 0.50 \\
                        & 0 & 2 &   0.15 & 0.15 & 0.30 &  0.20 & 0.30 \\
                        & 1 & 0 &   0.40 & 0.40 & 0.18 &  0.30 & 0.18\\
                        & 1 & 1 &   0.30 & 0.60 & 0.15 &  0.25 & 0.15\\
                        & 1 & 2 &   0.15 & 0.25 & 0.10 &  0.20 & 0.10\\
  \hline
   $\mu_s(a_1,a_2)$ & 0 & 0 & 0.603 & 0.603 & \bf{ 0.712} &  \bf{0.712} & \bf{0.658}\\
                         & 0 & 1 &   0.549  & 0.549 & 0.658 &  0.521  & \bf{0.658}\\
                        & 0 & 2 &   0.467  & 0.467 & 0.549 &  0.494 & 0.549\\
                        & 1 & 0 & \bf{0.660} & 0.603 & 0.557 &  0.613 & 0.557\\
                        & 1 & 1 &   0.613 & \bf{ 0.712} & 0.543 &  0.590 & 0.543 \\
                        & 1 & 2 &   0.541 & 0.521 & 0.520 &  0.566 & 0.520\\
  \hline
 \end{tabular}
\end{table}

In the simulations reported here, we took $n_s = 200$ aligned with the
sample size for the most prevalent subtype in I-SPY2.  To characterize
in-trial performance under each scenario, in Table~\ref{t:intrial}, we
show across the 5000 Monte Carlo trials under each randomization
scheme the Monte Carlo average overall pCR rate over all subjects; the
Monte Carlo average proportion of subjects whose in-trial experience
is consistent with having followed the optimal regime, with the
largest true pCR rate, and the worst regime, with the smallest; and
the Monte Carlo average of final randomization probabilities for the
stage 1 and 2 options associated with the optimal regime, the latter
for those who received the optimal stage 1 option.  Under Scenario 5,
we show the proportion of subjects with experience consistent with
either optimal regimes $\{0, 0\}$ and $\{0, 1\}$ and the average final
stage 2 randomization probabilities for both. Under the RAR
strategies, as expected, final randomization probabilities for the
treatment options associated with the optimal regime(s) are larger, in
some cases substantially so, than the nominal equal probabilities
under SR, with larger probabilities resulting from more aggressive
adaptation.  Under Scenario 5, proportions of subjects have experience
consistent with at least one of the optimal regimes are similar, as
are randomization probabilities to stage 2 options 0 and 1 among
subjects who received stage 1 option 0.

\begin{table}[ht]
 \centering
 \caption{In-trial results, $n_s = 200$. Overall PCR Rate is the Monte
   Carlo average pCR rate for subjects in the trial; Consist w/Opt is
   Monte Carlo average proportion of subjects with experience
   consistent with the optimal regime(s); Consist w/Worst is Monte Carlo
   average proportion of subjects with experience consistent with the
   worst regime; Rand Prob $a_1$ Opt is the Monte Carlo average final
   updated randomization probability for the stage 1 option $a_1$
   associated with the optimal regime; Rand Prob $a_2$ Opt is the
   Monte Carlo average final updated randomization probability for the
   stage 2 option $a_2$ associated with the optimal regime(s) for
   subjects randomized to the optimal stage 1 option.
   TS(\,$\cdot$\,) is defined in the text.}
 \label{t:intrial}
 \begin{small}
 \begin{tabular}{l c c c c c c c c} \Hline
Measure & Scenario & SR & TS(0.25) & TS(0.5) & TS(0.75) & TS(1) & TS($0.5t/T_{end}$)  & TS($t/T_{end}$) \\
\hline
Overall    & 1 &  0.572 & 0.580& 0.585 & 0.590 & 0.592 & 0.582& 0.587\\
pCR Rate & 2 &  0.575 & 0.588 & 0.600 & 0.606 & 0.610 & 0.593 & 0.604\\
               & 3 &  0.590 & 0.600 & 0.609 & 0.613 & 0.618 & 0.604 & 0.613\\
               & 4 &  0.582 & 0.594 & 0.603 & 0.609 &  0.614 & 0.598 & 0.607\\
               & 5 &  0.580 & 0.587 & 0.591 & 0.595 &  0.596 & 0.587 & 0.593\\*[0.03in]
   %\hline
Consist   & 1 &  0.258 & 0.288& 0.314 & 0.335 & 0.354 & 0.299 & 0.329\\
w/Opt     & 2 &  0.243 & 0.291 & 0.338 & 0.374 & 0.396 & 0.310 & 0.360\\
               & 3 &  0.244 & 0.286 & 0.330 & 0.359 & 0.377 & 0.305 & 0.351\\
               & 4 &  0.244 & 0.286 & 0.321  & 0.350& 0.365 & 0.304 & 0.343\\
               & 5 &  0.243 & 0.269 & 0.292  & 0.308 & 0.319 & 0.279 &  0.303\\
               &    &  0.243 & 0.268 & 0.291  & 0.309 & 0.322 & 0.280 & 0.306\\*[0.03in]   
 % \hline
Consist   & 1 & 0.243 & 0.212 & 0.191 & 0.176 & 0.167 & 0.205 & 0.184\\
w/Worst  & 2 & 0.243 & 0.210 & 0.186 & 0.172 & 0.163 & 0.201 & 0.179 \\
               & 3 & 0.257 & 0.230 & 0.208 & 0.193 & 0.185 & 0.222 & 0.197\\
               & 4 & 0.244 & 0.223 & 0.208 & 0.200&  0.194 & 0.216 & 0.203\\
               & 5 & 0.257 & 0.237 & 0.218 & 0.206&  0.196 & 0.228 & 0.209\\*[0.03in]
%\hline
Rand Prob & 1 & 0.500 & 0.563& 0.613 & 0.650 & 0.676 & 0.619& 0.676\\
$a_1$ Opt  & 2 & 0.500 & 0.589 & 0.667 & 0.720 & 0.750 & 0.661 & 0.746\\
                 & 3 & 0.500 & 0.622 & 0.716 & 0.776 & 0.810 & 0.715 & 0.816\\
                 & 4 & 0.500 & 0.539 & 0.575  & 0.614& 0.631 & 0.570 & 0.627\\
                 & 5 & 0.500 & 0.591 & 0.666  & 0.716& 0.631 & 0.664 & 0.752\\*[0.03in]
                 % \hline
Rand Prob & 1 & 0.333 & 0.409 & 0.463 & 0.505 &  0.536  & 0.471& 0.541\\
$a_2$ Opt  & 2 & 0.333 & 0.485 & 0.598 & 0.675 & 0.751 & 0.593 & 0.715\\
                 & 3 & 0.333 & 0.437 & 0.519 & 0.568 & 0.598 & 0.513  & 0.597\\
                 & 4 & 0.333 & 0.542 &  0.676 & 0.757&  0.794 & 0.682 & 0.810\\
                 & 5 & 0.333 & 0.380 &  0.410 & 0.425&  0.431 & 0.405 & 0.436\\
                 &   & 0.333 & 0.378 &  0.404 & 0.423&  0.444 & 0.412 & 0.441\\*[0.03in]
   \hline
 \end{tabular}
 \end{small}
\end{table}

\begin{figure}[h]
  \centering
  \includegraphics[width=2.25in]{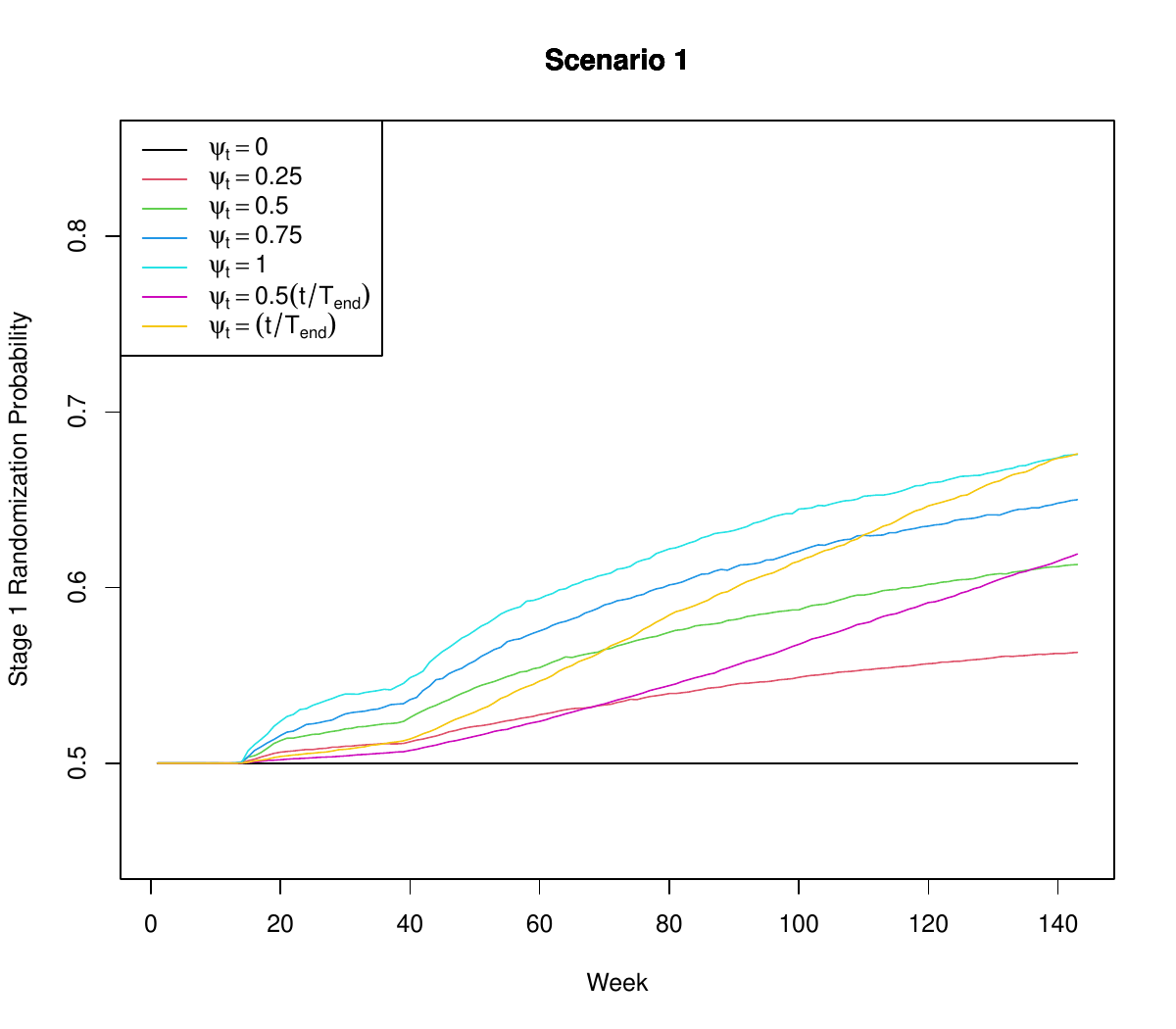}
  \includegraphics[width=2.25in]{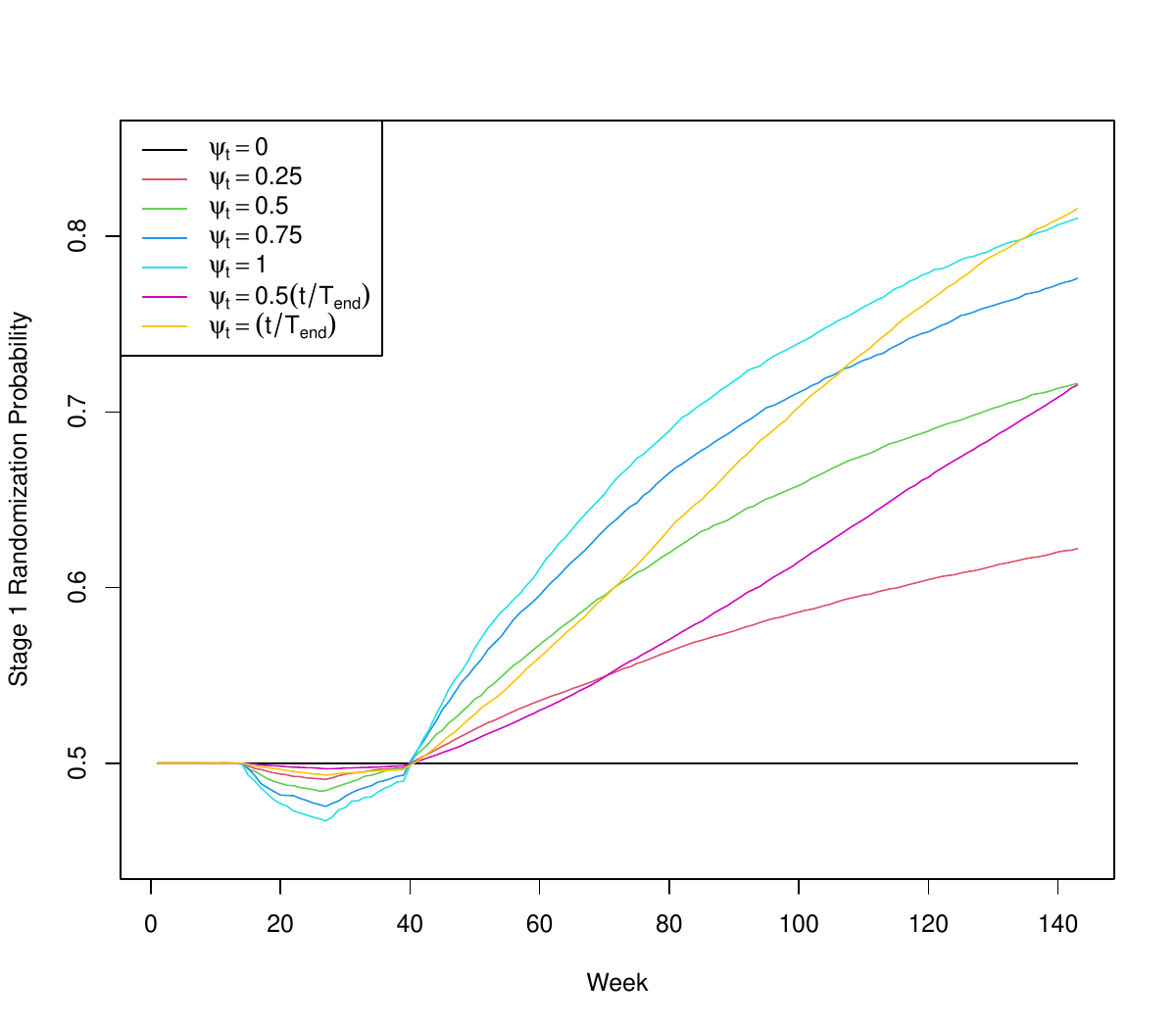}
\caption{Monte Carlo average randomization probabilities over the
  course of the trial to the stage 1 option associated with the
  optimal regime for Scenarios 1 and 3 under each randomization
  scheme.
\label{f:two}}
\end{figure}

Figure~\ref{f:two} presents for Scenarios 1 and 3 the Monte Carlo
average randomization probabilities to the stage 1 option associated
with the optimal regime under each randomization scheme.  Because in
antagonistic Scenario 3 option 0, that associated with the optimal
regime, has a lower pCR rate after stage 1 than option 1, following
the burn-in period, the probability of assignment to option 0 dips
below 0.5 for a short time until sufficient information accrues at
later stages to reflect its role in the optimal regime.  At that
point, the probabilities increase, with more dramatic increases
associated with more aggressive adaptive randomization. A similar
pattern arises for synergistic Scenario 4.  In constrast, under the no
delayed effects Scenario 1, option 1 has a higher pCR rate after stage
1 than option 0 and is also associated with the optimal regime; thus,
following the burn-in, randomization probabilities rise immediately to
favor this option.  Probabilities for Scenario 2 exhibit the same
behavior.

RAR results in increased proportions of subjects having in-trial
experience consistent with the optimal regime(s) relative to SR, with
higher proportions associated with more aggressive adaptation.  The
proportions consistent with the worst performing regime likewise
decrease relative to SR.
% , with lower proportions associated with more aggressive
% adaptation.
% In general, RAR leads to increased proportions with
% experience consistent with better-performing regimes (not shown).
These features are reflected in increases in the overall in-trial pCR
rate relative to that for SR.  Because this pCR rate is averaged over
results for all six embedded regimes and because RAR with clipping
constants implements exploration of all of them, the higher
probabilities of experience consistent with more efficacious regimes
can translate at best to modest increases in overall pCR rate shown in
Table~\ref{t:intrial}.

Table~\ref{t:posttrial} summarizes post-trial inference results for
Scenarios 1, 3, and 5; those for Scenarios 2 and 4 similar and are
presented in  Appendix D.  For each scenario, the Monte Carlo
averages of estimates of $\mu_s(a^{opt}_{1,s},a^{opt}_{2,s})$ obtained
using $\hatmu_{s,Bayes}(a_1,a_2)$, $\hatmu_{s,samp}(a_1,a_2)$, and
$\hatmu_{s,wtsamp}(a_1,a_2)$ are presented; in all cases, the Bayesian
estimator $\hatmu_{s,Bayes}(a_1,a_2)$ was calculated using $M=1000$.
Also shown for each estimator are Monte Carlo coverage of a 95\% Wald
confidence interval for $\mu_s(a^{opt}_{1,s},a^{opt}_{2,s})$, average
interval length, and the proportion of trials in which the estimator
correctly identifies the optimal regime.  For
$\hatmu_{s,samp}(a_1,a_2)$ and $\hatmu_{s,wtsamp}(a_1,a_2)$, Monte
Carlo efficiency relative to $\hatmu_{s,Bayes}(a_1,a_2)$, defined as
the Monte Carlo mean square error for $\hatmu_{s,Bayes}(a_1,a_2)$
divided by that for the given estimator, is presented.  For Scenario
5, we show results for regime $\{ 0, 0\}$; those for $\{0, 1\}$ are
similar.

\begin{table}
 \centering
 \caption{Post-trial results, $n_s = 200$.  For each estimator, Est
   pCR Rate Opt Regime is the Monte Carlo average of estimates of the
   value of the optimal regime; Coverage is the Monte Carlo coverage
   of a nominal 95\% Wald confidence interval for
   $\mu_s(a^{opt}_{1,s},a^{opt}_{2,s})$; Length is Monte Carlo average
   length of confidence intervals; 
   Prop Est Correct is the Monte
   Carlo proportion of trials for which the true optimal regime was
   correctly identified as optimal; and Rel Efficiency is the Monte
   Carlo relative efficiency of the indicated estimator to
   $\mu_{s,Bayes}(a_1,a_2)$.}
 \label{t:posttrial}
 \begin{footnotesize}
\begin{tabular}{l l c c c c c c c}     \Hline
Estimator & Measure & SR & TS(0.25) & TS(0.5) & TS(0.75) & TS(1) & TS($0.5t/T_{end}$)  & TS($t/T_{end}$) \\
   \hline
          & & \multicolumn{7}{c}{Scenario 1, $\mu_s(a^{opt}_{1,s},a^{opt}_{2,s})$ = 0.660} \\*[0.03in]
$\mu_{s,Bayes}(a_1,a_2)$ & Est pCR Rate Opt Regime$^{\rm a}$ & 0.640 & 0.636& 0.631 & 0.628& 0.626 & 0.635& 0.629\\
                                      & Coverage                     & 0.953 & 0.949& 0.942 & 0.937& 0.923 & 0.936&   0.935\\
                                        & Length                    & 0.274 & 0.260 & 0.252 & 0.249 & 0.247 & 0.255 &   0.249 \\
                                      & Prop Est Correct              & 0.487 & 0.501& 0.507 & 0.514& 0.524 & 0.517& 0.522\\*[0.03in]
$\mu_{s,samp}(a_1,a_2)$ & Est pCR Rate Opt Regime & 0.659 & 0.651& 0.644 & 0.639& 0.635 & 0.649& 0.640\\
                                      & Coverage                         & 0.939 & 0.939& 0.937 & 0.929& 0.927 & 0.935& 0.933\\
                                        & Length                        & 0.292&0.277&0.269&0.266&0.265&0.272&0.265\\
                                     &  Prop Est Correct              & 0.483 & 0.502& 0.506 & 0.511& 0.518 & 0.519& 0.519\\
                                      & Rel Efficiency                   & 0.825 & 0.898& 0.906 & 0.884& 0.861 & 0.902& 0.891\\*[0.03in]
   $\mu_{s,wtsamp}(a_1,a_2)$ & Est pCR Rate Opt Regime & --& 0.656& 0.651 & 0.648& 0.646 & 0.655& 0.649\\
                                      & Coverage                         & -- & 0.940& 0.939 & 0.934& 0.931 &0.934 & 0.935\\
                                        & Length                      & -- &0.277&0.270&0.270&0.270&0.272&0.267\\
                                      & Prop Est Correct              & -- & 0.502& 0.506 & 0.511& 0.518 & 0.519& 0.519\\
                                      &  Rel Efficiency                  & -- & 0.921& 0.969 & 0.960& 0.949 & 0.933& 0.968\\*[0.05in]

& & \multicolumn{7}{c}{Scenario 3, $\mu_s(a^{opt}_{1,s},a^{opt}_{2,s})$ = 0.712} \\*[0.03in]
$\mu_{s,Bayes}(a_1,a_2)$ & Est pCR Rate Opt Regime & 0.684 & 0.681 & 0.682 & 0.676& 0.675 & 0.681& 0.679\\
                                      & Coverage                         & 0.949 &0.939 & 0.946 &0.926 & 0.923 & 0.941& 0.929\\
                                        & Length                        & 0.266&0.244&0.233&0.226&0.226&0.237&0.226\\
                                     & Prop Est Correct              & 0.635 & 0.662& 0.692 & 0.684& 0.684 & 0.670&0.684 \\*[0.03in]
$\mu_{s,samp}(a_1,a_2)$ & Est pCR Rate Opt Regime & 0.712 & 0.704& 0.701 & 0.692& 0.690 & 0.702& 0.696\\
                                      & Coverage                         & 0.930 & 0.937& 0.945 & 0.935& 0.931 & 0.943& 0.933\\
                                        & Length                        & 0.279&0.254&0.243&0.237&0.237&0.247&0.235\\
                                      &  Prop Est Correct              & 0.648 & 0.668& 0.694 & 0.688& 0.684 & 0.678& 0.694\\
                                      & Rel Efficiency                   & 0.913 & 1.028& 1.069 & 1.013& 1.006 &1.059 & 1.047\\*[0.03in]
   $\mu_{s,wtsamp}(a_1,a_2)$ & Est pCR Rate Opt Regime & --& 0.707& 0.707 & 0.701& 0.700 & 0.707& 0.704\\
                                      & Coverage                         & -- & 0.936& 0.943 & 0.941& 0.937 & 0.939& 0.935\\
                                        & Length                      & --&0.253& 0.244&0.239&0.241&0.247&0.237\\
                                     & Prop Est Correct              & -- & 0.668& 0.694 & 0.688& 0.684 &0.678 & 0.694\\
                                      &  Rel Efficiency                  & -- & 1.059& 1.144 & 1.135& 1.136 &1.112 & 1.146\\*[0.05in]

& & \multicolumn{7}{c}{Scenario 5, $\mu_s(a^{opt}_{1,s},a^{opt}_{2,s})$ = 0.658} \\*[0.03in]
$\mu_{s,Bayes}(a_1,a_2)$ & Est pCR Rate Opt Regime & 0.637 & 0.632 &  0.627 &0.622 & 0.617 &  0.627& 0.623\\
                                      & Coverage                         & 0.951 & 0.934& 0.937 &0.930& 0.923 & 0.943& 0.936\\
                                        & Length                        & 0.277&0.266&0.261&0.262&0.264&0.264&0.261\\
                                      & Prop Est Correct              & 0.416 & 0.415& 0.423 & 0.424& 0.428 & 0.399& 0.428\\*[0.03in]
$\mu_{s,samp}(a_1,a_2)$ & Est pCR Rate Opt Regime & 0.658 & 0.649& 0.642 & 0.635& 0.628 & 0.643& 0.637\\
                                      & Coverage                         & 0.934 & 0.941& 0.935 & 0.931& 0.927& 0.941& 0.937\\
                                        & Length                        & 0.295&0.283&0.279&0.282&0.285&0.281&0.280\\
                                      &  Prop Est Correct              & 0.416 & 0.423& 0.432 & 0.432& 0.429 & 0.406& 0.432\\
                                      & Rel Efficiency                   & 0.837 & 0.914& 0.918 & 0.900& 0.860 & 0.946& 0.918\\*[0.03in]
   $\mu_{s,wtsamp}(a_1,a_2)$ & Est pCR Rate Opt Regime & --& 0.654& 0.650 & 0.645& 0.640 & 0.649& 0.646\\
                                      & Coverage                         & -- & 0.938& 0.937 & 0.931& 0.931 & 0.942& 0.939\\
                                        & Length                        & --&0.282&0.280&0.285&0.291&0.281&0.282\\
                                      & Prop Est Correct              & -- &0.423 & 0.432 & 0.432& 0.429 & 0.406& 0.432\\
                                      &  Rel Efficiency                  & -- & 0.939& 0.980 & 0.988& 0.943 & 0.983& 0.999\\*[0.05in]

   \hline
\end{tabular}
\end{footnotesize}
\end{table}

All estimators exhibit some downward bias in estimating the value of
the optimal regime(s) under all randomization schemes; the exception
is $\hatmu_{s,samp}(a_1,a_2)$ under SR, where standard asymptotic
theory holds.  The weighted estimator $\hatmu_{s,wtsamp}(a_1,a_2)$
shows the smallest bias overall.  Monte Carlo coverage probabilities
of confidence intervals in most cases are close to or slightly less
than the nominal 0.95 level, with lower coverage in many cases
associated with more aggressive adaptive randomization.  Overall,
length of confidence intervals based on $\hatmu_{s,Bayes}(a_1,a_2)$ is
shorter than that of those based on the frequentist estimators; that
for $\hatmu_{s,Bayes}(a_1,a_2)$ was shorter 80-95\% of the time across
all scenarios (not shown).  Under Scenarios 2-5, involving delayed
effects, the weighted estimator is most efficient for estimating the
value of the optimal regime and achieves only slightly less precision
than $\hatmu_{s,Bayes}(a_1,a_2)$ under Scenario 1, with no delayed
effects.  Under all scenarios, with SR, not surprisingly, the Bayesian
estimator yields improved precision over the sample proportion
estimator.

\begin{figure}[ht]
 \centering
\includegraphics[width=5.5in]{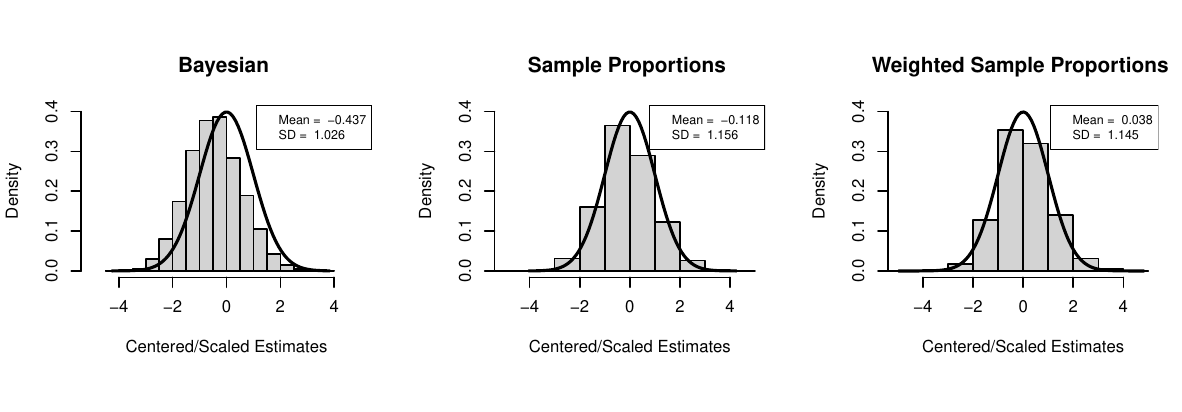}
\caption{Histograms of 5000 centered and scaled estimates of
  $\mu_s(a^{opt}_{1,s},a^{opt}_{2,s})$ for Scenario 3 under RAR with $\psi_t=1$
 using the
  Bayesian estimator $\hatmu_{s,Bayes}(a_1,a_2)$, estimator based on
  sample proportions $\hatmu_{s,samp}(a_1,a_2)$, and estimator based on
  weighted sample proportions $\hatmu_{s,wtsamp}(a_1,a_2)$. 
  \label{f:three}}
\end{figure}

Figure~\ref{f:three} presents histograms of the 5000 estimates of
$\mu_s(a^{opt}_{1,s},a^{opt}_{2,s})$ in Scenario 3 with TS(1) RAR
using all three estimators, centered by the true value and scaled by
standard error; ideally, these quantities should be approximately
standard normal.  Consistent with the findings of
\citet{ZhangMest2021} and \citet{NorwoodRAR} and with the bias noted
above, the distributions of $\hatmu_{s,Bayes}(a_1,a_2)$ and
$\hatmu_{s,samp}(a_1,a_2)$ are centered below zero, the latter
reflecting the failure of standard asymptotic theory under RAR.  The
histogram for $\hatmu_{s,wtsamp}(a_1,a_2)$ is centered at zero,
confirming that asymptotic normality is achieved through the
weighting.  Fortunately, the effects of this behavior on confidence
interval performance for the three estimators are unremarkable.
Finally, in all of Scenarios 1-5, the proportion of trials in which
each estimator identifies the optimal regime is increasingly higher
with increasing aggressiveness under RAR relative to SR.

 %%  could put this in  Appendix D
As noted in Section~\ref{s:intro}, a concern in conventional RCTs is
loss of efficiency of post-trial inference on treatment effects under
adaptive randomization relative to SR.  In  Appendix D, for
estimation of the value of the optimal regime under each scenario, we
present for each estimator Monte Carlo efficiency of the estimator
under each RAR scheme relative to under SR; because the natural
frequentist estimator under SR is $\hatmu_{s,samp}(a_1,a_2)$, that
shown for $\hatmu_{s,wtsamp}(a_1,a_2)$ is relative to
$\hatmu_{s,samp}(a_1,a_2)$ under SR.  The weighted estimator under RAR
is relatively more efficient than under SR for all but the most
aggressive adaptation; $\hatmu_{s,Bayes}(a_1,a_2)$ shows modest gains
or losses under less agressive randomization and moderate loss under
the most agressive schemes.  Overall, the results suggest concern over
efficiency loss is not great and justified mainly for the most
aggressive RAR schemes.

In Scenarios
1-4, % reflecting expectations of the investigators, there
there is a unique optimal regime, with some other regimes achieving
lower but similar pCR rates/values, while Scenario 5 involves two
regimes achieving the optimal value.  Our results demonstrate
satisfactory performance of $\hatmu_{s,Bayes}(a_1,a_2)$ and
$\hatmu_{s,wtsamp}(a_1,a_2)$ when there is more than one optimal
regime.  For further insight, in  Appendix D we consider an extreme
``null'' scenario in which all regimes achieve the same pCR rate.
Here, there is no effect of RAR on in-trial results, as expected; for
post-trial inference, $\hatmu_{s,Bayes}(a_1,a_2)$ achieves the best
overall performance.  See  Appendix D for more discussion.  These
results support the use of $\hatmu_{s,Bayes}(a_1,a_2)$ for post-trial
inference when several regimes are expected to be optimal or close to
optimal.

Although $n_s=200$ was chosen to reflect the conditions for I-SPY2, to
gauge the effect of sample size on performance, we also present in 
Appendix D additional simulation results for Scenario 3 under
$n_s = 120$ and 1000.  With $n_s = 1000$, the downward bias exhibited
by $\hatmu_{s,Bayes}(a_1,a_2)$, $\hatmu_{s,samp}(a_1,a_2)$, and, to a
lesser extent, $\hatmu_{s,wtsamp}(a_1,a_2)$ decreases, with that of
the latter negligible, suggesting that this feature is in part a
finite sample issue.

%
%  For post-trial inference, all estimators show downward
% bias; however, the Bayesian estimator achieves the best performance in
% terms of confidence interval coverage and efficiency relative to
% $\hatmu_{s,samp}(a_1,a_2)$ and $\hatmu_{s,wtsamp}(a_1,a_2)$ under all
% randomization schemes.  Moreover, efficiency of
% $\hatmu_{s,Bayes}(a_1,a_2)$ under adaptive randomization relative to
% SR is markedly better than that for the other two estimators.

Overall, our results suggest that, under the conditions expected in
the I-SPY2 SMART, the proposed RAR approach results in higher
proportions of trial participants being exposed to more efficacious
treatment strategies and modest improvements in overall pCR rate.  RAR
also results in improved ability to identify the optimal regime using
any of the estimators post trial, with the Bayesian and weighted
estimators achieving reliable performance.

\section{Discussion}\label{s:discuss}

We have reported on the Bayesian RAR strategy we developed for use in
the I-SPY2 SMART, which has been implemented in the ongoing trial.
In the I-ISPY2 SMART, the randomization probability updates are on a
fixed time schedule as in Section~\ref{s:adaptation}, providing a
standardized process for updating and review.  An important aspect of
review is to ensure that the implementation is aligning higher
randomization probabilities with more efficacious regimes;
probabilities and associated data are reviewed by the unblinded data
safety and monitoring board monthly as an additional safeguard.

% The strategy was developed for the I-SPY2 SMART, so the evaluation of
% performance reported here is under conditions similar to those
% expected in this trial.
A goal of transitioning I-SPY2 to a SMART was to provide participants
multiple opportunities to achieve pCR while receiving the minimum
amount of therapy needed to do so.  With the incorporation of RAR via
the proposed strategy, as demonstrated in our simulations, I-SPY2 will
assign more patients to more efficacious regimes, aligning with the
investigators’ objective of giving trial participants the greatest
chance of achieving pCR, while not compromising the quality of
post-trial inference for identifying the most efficacious treatment
strategies.  % Our simulations suggest that the proposed
% Bayesian and weighted estimators for subtype-specific mean regime
% outcome can be used reliably for post-trial inference.

Although we developed the proposed methods for I-SPY2, the general
approach to RAR based on Bayesian implementation of the g-computation
algorithm and Thompson sampling can be translated straightforwardly to
other SMART settings, as can the associated estimators for the values
of the embedded regimes, and we expect similar benefits would result.

\backmatter

\vspace*{-0.2in}

\section*{Acknowledgments}

This research was partially supported by National Institutes of Health
grants P01CA210961 (CY, DW, MD) and R01CA280970 (MD).

\bibliographystyle{biom} \bibliography{ispybib}

\begin{thebibliography}{}

\bibitem[\protect\citeauthoryear{Atkinson and Biswas}{Atkinson and
  Biswas}{2013}]{atkinson2013randomised}
Atkinson, A.~C. and Biswas, A. (2013).
\newblock {\em Randomised Response-Adaptive Designs in Clinical Trials}.
\newblock Chapman and Hall/CRC Press, Boca Raton, Florida.

\bibitem[\protect\citeauthoryear{Berry}{Berry}{2011}]{berry_promise}
Berry, D.~A. (2011).
\newblock Adaptive clinical trials: the promise and the caution.
\newblock {\em Journal of Clinical Oncology} {\bf 29,} 606--609.

\bibitem[\protect\citeauthoryear{Berry}{Berry}{2015}]{berry2015brave}
Berry, D.~A. (2015).
\newblock The brave new world of clinical cancer research: Adaptive
  biomarker-driven trials integrating clinical practice with clinical research.
\newblock {\em Molecular Oncology} {\bf 9,} 951--959.

\bibitem[\protect\citeauthoryear{Berry, Carlin, Lee, and Muller}{Berry
  et~al.}{2010}]{berry2010bayesian}
Berry, S.~M., Carlin, B.~P., Lee, J.~J., and Muller, P. (2010).
\newblock {\em Bayesian Adaptive Methods for Clinical Trials}.
\newblock Chapman and Hall/CRC Press, Boca Raton, Florida.

\bibitem[\protect\citeauthoryear{Cheung, Chakraborty, and Davidson}{Cheung
  et~al.}{2014}]{cheung_chakraborty_davidson_2014}
Cheung, Y.~K., Chakraborty, B., and Davidson, K.~W. (2014).
\newblock Sequential multiple assignment randomized trial {(SMART)} with
  adaptive randomization for quality improvement in depression treatment
  program.
\newblock {\em Biometrics} {\bf 71,} 450–459.

\bibitem[\protect\citeauthoryear{Das and Lo}{Das and Lo}{2017}]{ISPY2Das}
Das, S. and Lo, A. (2017).
\newblock Re-inventing drug development: A case study of the {I-SPY 2} breast
  cancer clinical trials program.
\newblock {\em Contemporary Clinical Trials} {\bf 62,} 168--174.

\bibitem[\protect\citeauthoryear{Hu and Rosenberger}{Hu and
  Rosenberger}{2006}]{hu2006theory}
Hu, F. and Rosenberger, W.~F. (2006).
\newblock {\em The Theory of Response-Adaptive Randomization in Clinical
  Trials}.
\newblock John Wiley and Sons, New York.

\bibitem[\protect\citeauthoryear{{I-SPY2~Trial~Consortium}}{{I-SPY2~Trial~Consortium}}{2020}]{ISPY2pCR}
{I-SPY2~Trial~Consortium} (2020).
\newblock Association of event-free and distant recurrence-free survival with
  individual-level pathological complete response in neoadjuvant treatment of
  stages 2 and 3 breast cancer: Three-year follow-up analysis for the {I-SPY2}
  adpatively randomized clinical trial.
\newblock {\em {JAMA} Oncology} {\bf 6,} 1355--1362.

\bibitem[\protect\citeauthoryear{Khoury, Meisel, Yau, Rugo, Nanda, Davidian,
  and {et al.}}{Khoury et~al.}{2024}]{ispy_smart2}
Khoury, K., Meisel, J.~L., Yau, C., Rugo, H.~S., Nanda, R., Davidian, M., and
  {et al.} (2024).
\newblock Datopotamab–deruxtecan in early-stage breast cancer: the sequential
  multiple assignment randomized {I-SPY2.2} phase 2 trial.
\newblock {\em Nature Medicine} {\bf 30,} 3737--3747.

\bibitem[\protect\citeauthoryear{Korn and Freidlin}{Korn and
  Freidlin}{2011}]{korn_freidlin_2011}
Korn, E.~L. and Freidlin, B. (2011).
\newblock Outcome-adaptive randomization: Is it useful?
\newblock {\em Journal of Clinical Oncology} {\bf 29,} 771–776.

\bibitem[\protect\citeauthoryear{Lavori and Dawson}{Lavori and
  Dawson}{2004}]{lavori_dawson}
Lavori, P.~W. and Dawson, R. (2004).
\newblock Dynamic treatment regimes: practical design considerations.
\newblock {\em Clinical Trials} {\bf 1,} 9--20.

\bibitem[\protect\citeauthoryear{Legocki, Meurer, Frederiksen, Lewis,
  Durkalski, Berry, Barsan, and Fetters}{Legocki
  et~al.}{2015}]{legocki_meurer_frederiksen_lewis_durkalski_berry_barsan_fetters_2015}
Legocki, L.~J., Meurer, W.~J., Frederiksen, S., Lewis, R.~J., Durkalski, V.~L.,
  Berry, D.~A., Barsan, W.~G., and Fetters, M.~D. (2015).
\newblock Clinical trialist perspectives on the ethics of adaptive clinical
  trials: A mixed-methods analysis.
\newblock {\em BMC Medical Ethics} {\bf 16,}.

\bibitem[\protect\citeauthoryear{Murphy}{Murphy}{2005}]{murphy_2005}
Murphy, S.~A. (2005).
\newblock An experimental design for the development of adaptive treatment
  strategies.
\newblock {\em Statistics in Medicine} {\bf 24,} 1455–1481.

\bibitem[\protect\citeauthoryear{Nanda, Liu, Yau, Shatsky, Puszati, Wallace,
  and {et al.}}{Nanda et~al.}{2020}]{jamapembro}
Nanda, R., Liu, M., Yau, C., Shatsky, R., Puszati, L., Wallace, A., and {et
  al.} (2020).
\newblock Effect of pembrolizumab plus neoadjuvant chemotherapy on pathologic
  complete response in women with early-stage breast cancer: An analysis of the
  ongoing phase 2 adaptively randomized {I-SPY2} trial.
\newblock {\em {JAMA} Oncology} {\bf 6,} 676--684.

\bibitem[\protect\citeauthoryear{Norwood, Davidian, and Laber}{Norwood
  et~al.}{2024}]{NorwoodRAR}
Norwood, P., Davidian, M., and Laber, E. (2024).
\newblock Adaptive randomization methods for sequential multiple assignment
  randomized trials ({SMART}s) via {Thompson} sampling.
\newblock {\em Biometrics} {\bf 80,} ujae152.

\bibitem[\protect\citeauthoryear{Shatsky, Trivedi, Yau, Nanda, Rugo, Davidian,
  and {et al.}}{Shatsky et~al.}{2024}]{ispy_smart}
Shatsky, R.~A., Trivedi, M.~A., Yau, C., Nanda, R., Rugo, H., Davidian, M., and
  {et al.} (2024).
\newblock Datopotamab–deruxtecan plus durvalumab in early-stage breast
  cancer: the sequential multiple assignment randomized {I-SPY2.2} phase 2
  trial.
\newblock {\em Nature Medicine} {\bf 30,} 3728--3736.

\bibitem[\protect\citeauthoryear{Thall and Wathen}{Thall and
  Wathen}{2007}]{practical_thompson}
Thall, P. and Wathen, K. (2007).
\newblock Practical {Bayesian} adaptive randomization in clinical trials.
\newblock {\em European Journal of Cancer} {\bf 43,} 859--866.

\bibitem[\protect\citeauthoryear{Thall, Fox, and Wathen}{Thall
  et~al.}{2015}]{thall_caveats}
Thall, P.~F., Fox, P.~S., and Wathen, J.~K. (2015).
\newblock Some caveats for outcome adaptive randomization in clinical trials.
\newblock In Sverdiov, O., editor, {\em Modern Adaptive Randomized Clinical
  Trials: Statistical and Practical Aspects.\,\,\,}, pages 287--305, Boca
  Raton, Florida. Chapman and Hall/CRC Press.

\bibitem[\protect\citeauthoryear{Thompson}{Thompson}{1933}]{thompson1933likelihood}
Thompson, W.~R. (1933).
\newblock On the likelihood that one unknown probability exceeds another in
  view of the evidence of two samples.
\newblock {\em Biometrika} {\bf 25,} 285--294.

\bibitem[\protect\citeauthoryear{Tsiatis, Davidian, Holloway, and
  Laber}{Tsiatis et~al.}{2020}]{tsiatis2020dynamic}
Tsiatis, A.~A., Davidian, M., Holloway, S., and Laber, E. (2020).
\newblock {\em Dynamic Treatment Regimes: Statistical Methods for Precision
  Medicine}.
\newblock Chapman and Hall/CRC Press, Boca Raton, Florida.

\bibitem[\protect\citeauthoryear{Wang, Wu, and Wahed}{Wang
  et~al.}{2022}]{wang_wu_wahed_2021}
Wang, J., Wu, L., and Wahed, A.~S. (2022).
\newblock Adaptive randomization in a two-stage sequential multiple assignment
  randomized trial.
\newblock {\em Biostatistics} {\bf 23,} 1182--1199.

\bibitem[\protect\citeauthoryear{Wathen and Thall}{Wathen and
  Thall}{2017}]{thall_wathen_2017}
Wathen, J.~K. and Thall, P.~F. (2017).
\newblock A simulation study of outcome adaptive randomization in multi-arm
  clinical trials.
\newblock {\em Clinical Trials} {\bf 14,} 432--440.

\bibitem[\protect\citeauthoryear{Yang, Cheng, Thall, and Wahed}{Yang
  et~al.}{2024}]{YangThallWahed}
Yang, X., Cheng, Y., Thall, P.~F., and Wahed, A.~S. (2024).
\newblock A generalized outcome-adaptive sequential multiple assignment
  randomized trial design.
\newblock {\em Biometrics} {\bf 80,} ujae073.

\bibitem[\protect\citeauthoryear{Zhang, Janson, and Murphy}{Zhang
  et~al.}{2021a}]{zhang2021inference}
Zhang, K.~W., Janson, L., and Murphy, S.~A. (2021a).
\newblock Inference for batched bandits.
\newblock {\em arXiv preprint arXiv:2002.03217} .

\bibitem[\protect\citeauthoryear{Zhang, Janson, and Murphy}{Zhang
  et~al.}{2021b}]{ZhangMest2021}
Zhang, K.~W., Janson, L., and Murphy, S.~A. (2021b).
\newblock Statistical inference with {M}-estimators on adaptively collected
  data.
\newblock {\em Advances in Neural Information Processing Systems} {\bf 34,}
  7460--7471.

\end{thebibliography}

\clearpage

\appendix

\setcounter{equation}{0}
\renewcommand{\theequation}{A.\arabic{equation}}

\section*{A. Summary of notation}\label{s:notation}

\begin{table}
 \centering
\caption{Summary of notation, subjects previously enrolled by week $t$.}
\label{t:one}
\begin{small}
\begin{tabular}{l l}   \Hline
Symbol & Definition \\
  \hline
 \multicolumn{2}{l}{Stage 1}  \\*[0.01in]
$n_{1,s,t}(a_1)$ & Number of subjects from subtype $s$ with $A_1=a_1$ who have $R_1$ observed before week $t$ \\
$R^{+}_{1,s,t}(a_1)$ & Number of subjects from subtype $s$ with $A_1=a_1$ who have $R_1=1$ before week $t$ \\
$n_{1,s,t}^{*}(a_1)$ & Number of subjects from subtype $s$ with $A_1=a_1$, $R_1=1$ who have $Y_1$ observed before week $t$ \\
$Y^{+}_{1,s,t}(a_1)$ & Number of subjects from subtype $s$ with $A_1=a_1$, $R_1=1$ who have $Y_1=1$ before week $t$ \\*[0.15in]

 \multicolumn{2}{l}{Stage 2 $(R_1 = 0)$}  \\*[0.01in]
$n_{2,s,t}(a_1,a_2)$ & Number of subjects from subtype $s$ with $(A_1,A_2)=(a_1,a_2)$ who have $R_2$ observed before week $t$. \\
$R^{+}_{2,s,t}(a_1,a_2)$ & Number of subjects from subtype $s$ with $(A_1,A_2)=(a_1,a_2)$ who have $R_2=1$ before week $t$ \\
$n_{2,s,t}^{*}(a_1,a_2)$ & Number of subjects from subtype $s$ with $(A_1,A_2)=(a_1,a_2)$, $R_2=1$ who have $Y_2$ observed before week $t$ \\
$Y^{+}_{2,s,t}(a_1,a_2)$ & Number of subjects from subtype $s$ with $(A_1,A_2)=(a_1,a_2)$, $R_2=1$ who have  $Y_2=1$ before week $t$ \\*[0.15in]

 \multicolumn{2}{l}{Stage 3 $(R_1 = 0, R_2=0)$}   \\*[0.01in]
$n_{3,s,t}^{*}(a_1,a_2)$ & Number of subjects from subtype $s$ with $(A_1,A_2)=(a_1,a_2)$ who have $Y_3$ observed before week $t$ \\
$Y^{+}_{3,s,t}(a_1,a_2)$ & Number of subjects from subtype $s$ with $(A_1,A_2)=(a_1,a_2)$ who have $Y_3=1$ before week $t$ \\
  \hline
\end{tabular}
\end{small}
\end{table}

\setcounter{equation}{0}
\renewcommand{\theequation}{B.\arabic{equation}}

\section*{B. Formulation of value estimators}\label{s:value}

For convenience, we repeat the following definitions from the main
paper.  For subtype $s$ and $a_1 \in \mathcal{A}_{1s}$ and
$a_2 \in \mathcal{A}_{2s}$,
\begin{equation}
\theta_1(a_1) = P(R_1 = 1 \mid S=s, A_1=a_1), \hspace{0.1in}
\gamma_1(a_1) = P(Y_1 =1 \mid S=s, A_1=a_1,R_1=1),
\label{eq:thetagamma1.supp}
\end{equation}
\vspace*{-0.35in}
\begin{equation}
\label{eq:thetagamma2.supp}
  \begin{aligned}
&\theta_2(a_1,a_2) = P(R_2= 1 \mid S=s, A_1=a_1,R_1=0, A_2=a_2),\\
\hspace{0.07in}
\gamma_2(a_1,&a_2) = P(Y_2 = 1 \mid S=s, A_1=a_1, R_1=0, A_2=a_2,
R_2=1).
\end{aligned}
\end{equation}
and
\begin{equation}
\gamma_{3,s}(a_1,a_2) = P(Y_3=1 \mid S=s, A_1=a_1, R_1=0, A_2=a_2,
R_2=0).
\label{eq:gamma3.supp}
\end{equation}
Then the value of the subtype $s$-specific regime $\{a_1,a_2\}$, i.e.,
the probability of achieving pCR if the population of subjects of
subtype $s$ were to follow regime $\{a_1,a_2\}$, is 
\begin{equation}
   \label{eq:value.supp} 
\begin{aligned}
 \mu_s(a_1,a_2) &= \theta_1(a_1) \gamma_1(a_1) +
 \{1-\theta_1(a_1)\} \theta_2(a_1,a_2) \gamma_2(a_1,a_2) \\
&+ \{1-\theta_1(a_1)\}\{1-\theta_2(a_1,a_2)\}
\gamma_{3,s}(a_1,a_2). 
\end{aligned}
\end{equation}

Suppose that there are $n$ subjects across all subtypes, $n_s$ of whom
are from of subtype $s$.  Note that at any week $t$, as seen in Table
1 in the main paper, the part of the available data $\mathcal{D}_t$ on
previously enrolled subjects that are relevant to updating
randomization probabilities for subjects of subtype $s$ at week $t$
are the subsets of $\mathcal{O}$ that have already been observed
before $t$ on among the $n_s$ subjects who enrolled prior to $t$.
Indexing these $n_s$ subjects by $i$, at the conclusion of the trial,
the part of the final data $\mathcal{D}_{final}$ relevant to
estimating $\mu_s(a_1, a_2)$ for subtype $s$ are $\mathcal{O}_i$,
$i = 1,\ldots,n_s$.  As discussed in the main paper, the proposed
Bayesian estimator $\hatmu_{s,Bayes}(a_1,a_2)$ is calculated based on
these data.  Namely, with 
$$\Theta_{1,s} = \{ \gamma_{1,s}(a_1), \gamma_{2,s}(a_1,a_2),
\gamma_{3,s}(a_1,a_2), \theta_{1,s}(a_1), \theta_{2,s}(a_1,a_2);\,\,\,
(a_1,a_2) \in \mathcal{A}_{1,s} \times \mathcal{A}_{2,s} \},$$ to
estimate the value of the subtype $s$-specific embedded regime
$\{a_1,a_2\}$, draw a sample of size $M$ from each of the posterior
distributions given $\mathcal{D}_{final}$ of the components of
$\Theta_{1,s}$ to obtain draws $\Theta_{1,s}^{(m)}$, $m=1,\ldots,M$,
from the joint posterior of $\Theta_{1,s}$ given
$\mathcal{D}_{final}$.  Then substitute each of these draws in
(\ref{eq:value.supp}) to obtain $\mu^{(m)}_s(a_1,a_2)$, $m = 1,\ldots,M$,
which can be viewed as a sample from the posterior distribution of
$\mu_s(a_1, a_2)$ given $\mathcal{D}_{final}$.  The Bayesian estimator
$\hatmu_{s,Bayes}(a_1,a_2)$ is obtained as the mean or mode of the
sample, with the standard deviation of the sample as a measure of
uncertainty.

We now present the formulation of the alternative estimators
$\hatmu_{s,samp}(a_1,a_2)$ and $\hatmu_{s,wtsamp}(a_1,a_2)$ based on
the final data $\mathcal{O}_i$, $i = 1,\ldots,n_s$ on the subjects of
subtype $s$.  These estimators may have appeal in that they can be
considered as frequentist alternatives to $\hatmu_{s,Bayes}(a_1,a_2)$.
First, note from (\ref{eq:thetagamma1.supp})-(\ref{eq:gamma3.supp}) that
$$\theta_1(a_1) \gamma_1(a_1) = \frac{\alpha_{1,s}(a_1)}{\alpha_{2,s}(a_1)} =
\frac{P(S=s, A_1=a_1, R_1=1, Y_1=1)}{P(S=s, A_1=a_1)},$$
$$\theta_2(a_1,a_2) \gamma_2(a_1,a_2) =
\frac{\alpha_{3,s}(a_1,a_2)}{\alpha_{4,s}(a_1,a_2)} = \frac{P(S=s, A_1=a_1, R_1=0, A_2=a_2,
  R_2=1, Y_2=1)}{P(S=s, A_1=a_1, R_1=0, A_2=a_2)},$$
$$\{1-\theta_2(a_1,a_2)\} \gamma_2(a_1,a_2) =
\frac{\alpha_{5,s}(a_1,a_2)}{\alpha_{4,s}(a_1,a_2)} =
\frac{P(S=s, A_1=a_1, R_1=0, A_2=a_2,
  R_2=0, Y_3=1)}{P(S=s, A_1=a_1, R_1=0, A_2=a_2)},$$
$$\{1-\theta_1(a_1)\} = \frac{\alpha_{6,s}(a_1)}{\alpha_{2,s}(a_1)} =
\frac{P(S=s, A_1=a_1, R_1=0)}{P(S=s, A_1=a_1)}.$$
Thus, $\mu_s(a_1, a_2)$ also can be written as
\begin{equation}
\mu_s(a_1, a_2) = \frac{\alpha_1(a_{1,s})}{\alpha_{2,s}(a_1)} +
\frac{\alpha_{6,s}(a_1)}{\alpha_{2,s}(a_1)} \left\{ \frac{\alpha_{3,s}(a_1,a_2)+\alpha_{5,s}(a_1,a_2)}{\alpha_{4,s}(a_1,a_2)}\right\}.
  \label{eq:alphavalue}
  \end{equation}
  Given the representation in (\ref{eq:alphavalue}), a natural plug-in
  estimator is based on substituting sample proportions for the
 probabilities $\alpha_{1,s}(a_1), \alpha_{2,s}(a_1),
 \alpha_{3,s}(a_1,a_2), \alpha_{4,s}(a_1,a_2), \alpha_{5,s}(a_1,a_2),
 \alpha_{6,s}(a_1)$, namely, 
  \begin{equation}
\hatmu_{s,samp}(a_1, a_2) = \frac{\hatalpha_{1,s}(a_1)}{\hatalpha_{2,s}(a_1)} +
\frac{\hatalpha_{6,s}(a_1)}{\hatalpha_{2,s}(a_1)} \left\{ \frac{\hatalpha_{3,s}(a_1,a_2)+\hatalpha_{5,s}(a_1,a_2)}{\hatalpha_{4,s}(a_1,a_2)}\right\},
\label{eq:musamp}    
\end{equation}
where
$$\hatalpha_{1,s}(a_1) =  n^{-1}_s\sumins I(A_{1i}=a_1, R_{1i}=1, Y_{1i}=1),
\hspace{0.1in}
\hatalpha_{2,s}(a_1) = n^{-1}_s\sumins I(A_{1i}=a_1),$$
$$\hatalpha_{3,s}(a_1,a_2) = n^{-1}_s\sumins I(A_{1i}=a_1, R_{1i}=0, A_{2i}=a_2,
R_{2i}=1, Y_{2i}=1),$$
$$\hatalpha_{4,s}(a_1,a_2) = n^{-1}_s\sumins I(A_{1i}=a_1,
R_{1i}=0, A_{2i}=a_2),$$
$$\hatalpha_{5,s}(a_1,a_2) =  n^{-1}_s\sumins I(A_{1i}=a_1, R_{1i}=0, A_{2i}=a_2, R_{2i}=0,
Y_{3i}=1),$$
$$\hatalpha_{6,s}(a_1) = n^{-1}_s\sumins
I(A_{1i}=a_1, R_{1i}=0).$$ Note that in these expressions we do not
include $S_i = s$ in the indicator functions, as the sums are over
only subjects $i$ for whom $S_i = s$ is true.

As discussed in Section 3.3 of the main paper, because of the use of
adaptive randomization, the final data from subjects of subtype $s$,
$\mathcal{O}_i$, $i = 1,\ldots,n_s$, are not independent and
identically distributed (i.i.d.) over $i$, so that standard asymptotic
theory may not apply to derive the properties of (\ref{eq:musamp}).
Thus, it is not even clear that the estimators above are consistent
estimators for the corresponding probabilities.  Nonetheless,
proceeding naively as if the data are i.i.d. and applying standard
theory, one can derive an approximate standard error for
$\hatmu_{s,samp}(a_1, a_2)$ by finding the associated influence
function.  We provide a heuristic argument.  Denote by
$\as_{1,s}(a_1), \as_{2,s}(a_1), \as_{3,s}(a_1, a_2), \as_{4,s}(a_1,
a_2), \as_{5,s}(a_1, a_2), \as_{6,s}(a_1)$ the limits in probability
of
$\hatalpha_1(a_1), \hatalpha_2(a_1), \hatalpha_{3,s}(a_1 ,a_2),
\hatalpha_{4,s}(a_1 ,a_2), \hatalpha_{5,s}(a_1 ,a_2),
\hatalpha_{6,s}(a_1)$, which would be the corresponding true probabilities
if the data were i.i.d.  Let
$$\mu^*_s(a_1, a_2) = \frac{\as_{1,s}(a_1)}{\as_{2,s}(a_1)} +
\frac{\as_{6,s}(a_1)}{\as_{2,s}(a_1)}
\left\{  \frac{\as_{3,s}(a_1,a_2)+\as_{5,s}(a_1,a_2)}{\as_{4,s}(a_1,a_2)} \right\}.$$
We find the mean-zero quantity $\mathcal{I}_i$ such that
\begin{equation}
n^{1/2}_s \left\{ \hatmu_s(a_1, a_2) - \mu^*_s(a_1, a_2)\right\} = n^{-1/2}_s
\sumins \mathcal{I}_i + o_p(1).
\label{eq:influence}
\end{equation}
As is well known, with i.i.d. data, it follows that the left hand side
of (\ref{eq:influence}) converges in distribution to a normal random
variable with mean zero and variance that can be approximated by
$n^{-1}_s \sumins \mathcal{I}_i^2$.  Assuming that
$n^{1/2}_s \{\hatalpha_{j,s}(a_1)-\as_{j,s}(a_1)\} = O_p(1)$, $j=1, 2, 6$, and
$n^{1/2}_s\{\hatalpha_{j,s}(a_1,a_2)-\as_{j,s}(a_1,a_2)\} = O_p(1)$, $j=3, 4, 5$,
so are bounded in probability, it is straightforward by tedious
algebra to show that
\begin{equation}
  \label{eq:sampmu}
  \begin{aligned}
&n^{1/2}_s\left\{ \hatmu_{s,samp}(a_1,a_2) - \mu^*_s(a_1, a_2)\right\}\\
&= \left\{\frac{1}{\as_{2,s}(a_1)} \right\}\left( n^{1/2}_s\{ \hatalpha_{1,s}(a_1)-\as_{1,s}(a_1)\} - \mu^*_s(a_1, a_2)\,
  n^{1/2}_s\{\hatalpha_{2,s}(a_1)-\as_{2,s}(a_1)\} \vphantom{\frac{n^{1/2}_s(\hatalpha_{4,s}(a_1,a_2)-\as_{4,s}(a_1,a_2))}{\as_{4,s}(a_1,a_2)}}\right.\\
&\hspace{0.05in}+\left\{\frac{\as_{3,s}(a_1,a_2)+\as_{5,s}(a_1,a_2)}{\as_{4,s}(a_1,a_2)}\right\}
\\
&\hspace{0.2in}\times\left[ -\left\{\frac{\as_{6,s}(a_1)}{\as_{4,s}(a_1,a_2)}\right\}
  n^{1/2}_s\{\hatalpha_{4,s}(a_1,a_2)-\as_{4,s}(a_1,a_2)\} + n^{1/2}_s\{\hatalpha_{6,s}(a_1)-\as_{6,s}(a_1)\}\right]\\
&\hspace{0.05in}+\left\{ \frac{\as_{6,s}(a_1)}{\as_{4,s}(a_1,a_2)} \right\} n^{1/2}_s\left[\{
  \hatalpha_{3,s}(a_1,a_2)+\hatalpha_{5,s}(a_1,a_2)\}-\{\as_{3,s}(a_1,a_2)+\as_{5,s}(a_1,a_2)\}\right]
\left.\vphantom{\frac{n^{1/2}_s(\hatalpha_{4,s}(a_1,a_2)-\as_{4,s}(a_1,a_2))}{\as_{4,s}(a_1,a_2)}}\right) + o_p(1).
  \end{aligned}
  \end{equation}
  Substituting the expressions for
  $\hatalpha_{1,s}(a_1), \hatalpha_{2,s}(a_1), \hatalpha_{3,s}(a_1,a_2), \hatalpha_{4,s}(a_1,a_2),
  \hatalpha_{5,s}(a_1,a_2), \hatalpha_{6,s}(a_1)$ in (\ref{eq:sampmu}) and simplifying leads to
\begin{align}
\mathcal{I}_i &= \left\{ \frac{1}{\as_{2,s}(a_1)}\right\} \left( I(A_{1i}=a_1, R_{1i}=1, Y_{1i}=1) - \as_{1,s}(a_1) -
  \mu^*_s(a_1, a_2)\{ I(A_{1i}=a_1) - \as_{2,s}(a_1)\} \vphantom{\left(\frac{\as_{3,s}(a_1,a_2)+\as_{5,s}(a_1,a_2)}{\as_{4,s}(a_1,a_2)}\right)}\right. \nonumber \\
  &\hspace{0.05in}+ \left\{
    \frac{\as_{3,s}(a_1,a_2)+\as_{5,s}(a_1,a_2)}{\as_{4,s}(a_1,a_2)}\right\}
    \left[ -\left\{\frac{\as_{6,s}(a_1)}{\as_{4,s}(a_1,a_2)}\right\} \right.
    \nonumber \\
  &\hspace*{0.1in}\times \left.\left\{ I(A_{1i}=a_1,R_{1i}=0,
      A_{2i}=a_2) - \as_{4,s}(a_1,a_2)\right\} +
    \left\{I(A_{1i}=a_1,R_{1i}=0) - \as_{6,s}(a_1)\right\} \vphantom{\frac{\as_{3,s}(a_1,a_2)+\as_{5,s}(a_1,a_2)}{\as_{4,s}(a_1,a_2)}} \right] \nonumber \\
  &\hspace{0.05in}+ \left\{ \frac{\as_{6,s}(a_1)}{\as_{4,s}(a_1,a_2)}\right\}
    \left[ \vphantom{ \{\as_{3,s}(a_1,a_2)+\as_{5,s}(a_1,a_2\} }
    I(A_{1i}=a_1, R_{1i}=0, A_{2i}=a_2, R_{2i}=1,Y_{2i}=1)\right. \nonumber\\
&\hspace{0.05in}  +   I(A_{1i}=a_1, R_{1i}=0, A_{2i}=a_2, R_{2i}=0, Y_{3i}=1)
    \nonumber \\
&\hspace*{0.85in} \left. \left. - \{\as_{3,s}(a_1,a_2)+\as_{5,s}(a_1,a_2\}
    \vphantom{ \left[  I(A_{1i}=a_1, R_{1i}=0, A_{2i}=a_2,
        R_{2i}=1,Y_{2i}=1) + I(A_{1i}=a_1, R_{1i}=0, A_{2i}=a_2,
        R_{2i}=0, Y_{3i}=1)\right]}  \right] \vphantom{ \left\{
      \frac{\as_{6,s}(a_1)}{\as_{4,s}(a_1,a_2)}\right\} }    \right).
  \label{eq:if}
\end{align}
Assuming that $n_s^{-1} \sumins \mathcal{I}_i^2$ converges in
probability to $\Sigma > 0$, we can conclude that
$$n_s^{1/2}\left\{ \hatmu_{s,samp}(a_1,a_2) -
  \mu^*_s(a_1, a_2)\right\}$$ converges in distribution to a mean-zero
normal random variable with variance $\Sigma$.  

For practical use, letting $\widehat{\mathcal{I}}_i$ denote
(\ref{eq:if}) with
$\hatalpha_{1,s}(a_1), \hatalpha_{2,s}(a_1), \hatalpha_{3,s}(a_1,a_2),
\hatalpha_{4,s}(a_1,a_2), \hatalpha_{5,s}(a_1,a_2)$,
$\hatalpha_{6,s}(a_1)$ and $\hatmu_{s,samp}(a_1, a_2)$ substituted for
$\as_{1,s}(a_1), \as_{2,s}(a_1), \as_{3,s}(a_1, a_2)$, $\as_{4,s}(a_1,
a_2), \as_{5,s}(a_1, a_2), \as_{6,s}(a_1)$ and $\mu^*_s(a_1, a_2)$,
estimate $\Sigma$ by $n^{-1}_s \sumins \widehat{\mathcal{I}}_i^2$.  

The foregoing result was derived by taking $\mathcal{O}$,
$i=1,\ldots, n_s$, to be i.i.d. and applying standard theory.  Because
these data are not i.i.d. owing to the adaptive randomization,
inferences based on this result are likely to be flawed.  Accordingly,
we now consider derivation of the estimator
$\hatmu_{s,wtsamp}(a_1,a_2)$, which, as discussed in Section 3.3 of
the main paper, involves weighted versions of estimators for each of
the probabilities in (\ref{eq:alphavalue}).  The derivation is
heuristic and follows the spirit of arguments in \citet{NorwoodRAR},
which adapts the approach of \citet{ZhangMest2021} on theory for
M-estimators based on adaptively collected data.

Define the set of all potential outcomes for a randomly chosen patient
in the subtype $s$ population as
$$\mathcal{W}^* = \{R_1^*(a_1), Y_1^*(a_1), R_2^*(a_1,a_2), 
Y_2^*(a_1,a_2), Y_3^*(a_1,a_2), \hspace{0.05in} \mbox{for all}
\hspace{0.05in} (a_1,a_2) \in \mathcal{A}_{1,s} \times
\mathcal{A}_{2,s} \}.$$ Here, $R_1^*(a_1)$ is the response status and
$Y_1^*(a_1)$ is the pCR status a patient would achieve if given stage
1 treatment $a_1$; $R_2^*(a_1, a_2)$ is the response status a patient
would achieve if given stage 1 treatment $a_1$ followed by stage 2
treatment $a_2$; and $Y_2^*(a_1,a_2)$ is the pCR status a patient
would achieve if given stage 1 treatment $a_1$ followed by stage 2
treatment $a_2$ at the end of stage 2, and $Y_32^*(a_1,a_2)$ is the
pCR status at the end of stage 3.  Although the observed data
$\mathcal{O}_i$, $i = 1,\ldots,n_s$, are not i.i.d., the potential
outcomes $\mathcal{W}^*_i$, $i = 1,\ldots,n_s$, are i.i.d.  We make
standard identifiability assumptions, which are discussed extensively
elsewhere \citep{tsiatis2020dynamic}: (i) consistency, which implies
that observed variables are equal to their potential counterparts
under the treatments actually received; (ii) positivity, which implies
that the probabilities of being assigned any treatment in
$\mathcal{A}_{1,s}$ and $\mathcal{A}_{2,s}$ are strictly greater than
zero, which holds in ISPY2.2 through the use of ``clipping constants''
that prevent adaptive probabilities from dropping below a set
threshold; and (iii) sequential randomization, which holds by design
in a nonadaptively randomized SMART and is discussed further below in
the case the adaptively randomized ISPY2.2 SMART.

From (\ref{eq:alphavalue}), we can write
\begin{equation}
\mu_s(a_1, a_2) = \phi_{1,s}(a_1) + \phi_{2,s}(a_1) \{ \phi_{3,s}(a_1,a_2) + \phi_{4,s}(a_1,a_2)\},
\label{eq:phivalue}
\end{equation}
where 
\begin{align}
 \phi_{1,s}(a_1) &= \frac{\alpha_1(a_{1,s})}{\alpha_{2,s}(a_1} = P(R_1=1, Y_1=1 \mid S=s, A_1=a_1) \label{eq:phi1} \\
 \phi_{2,s}(a_1) &= \frac{\alpha_{6,s}(a_1)}{\alpha_2(a_1)}  = P(R_1 =
 0 \mid S=s, A_1=a_1) \label{eq:phi2} \\
 \phi_{3,s}(a_1,a_2) &=
 \frac{\alpha_{3,s}(a_1,a_2)}{\alpha_{4,s}(a_1,a_2)} = P(R_2=1, Y_2=1
   \mid S=s, A_1=a_1,R_1=0,A_2=a_2) \label{eq:phi3} \\
  \phi_{4,s}(a_1,a_2) &=
 \frac{\alpha_{5,s}(a_1,a_2)}{\alpha_{4,s}(a_1,a_2)} = P(R_2=0, Y_3=1 
   \mid S=s, A_1=a_1,R_1=0,A_2=a_2) \label{eq:phi4} 
\end{align}
For some set of weights $\wone_i(a_1)$, $i = 1,\ldots,n_s$, where we
reiterate that $i$ indexes only subjects with $S=s$, define
$$\hatalpha^w_{1,s}(a_1) =  n^{-1}_s\sumins \wone_i(a_1)  I(A_{1i}=a_1, R_{1i}=1, Y_{1i}=1),
\hspace{0.15in} \hatalpha^w_{2,s}(a_1) = n^{-1}_s \sumins \wone_i(a_1) 
I(A_{1i}=a_1).$$ Then note that the obvious estimator for
$\phi_{1,s}(a_1)$ in (\ref{eq:phi1}) is
$\hatphi_{1,s}(a_1) = \hatalpha^w_{1,s}(a_1)/\hatalpha^w_{2,s}(a_1)$,
which solves in $\phi_{1,s}(a_1)$ an M-estimating equation of the form
\begin{equation}
\sumins \wone_i(a_1)  I(A_{1i}=a_1)\{ I(R_{1i}=1, Y_{1i}=1) -
\phi_{1,s}(a_1)\} = \sumins M\{ A_{1i}, R_{1i}, Y_{1i}, a_1; \phi_{1,s}(a_1)\} = 0.
\label{eq:Mest1}
\end{equation}
Following \citet{ZhangMest2021} and Norwood et
al. (2024) \citet{NorwoodRAR}, we choose the $\wone_i(a_1)$ so that the
estimator $\hatphi_{1,s}(a_1)$ is consistent for $\phi_{1,s}(a_1)$
with asymptotic normal theory that can be established via the
martingale central limit theorem.  Assume for definiteness that
subject $i$ enrolled during the interval $[t, t+1)$, so was assigned
to stage 1 treatment according to randomization probabilities
$\pi_{1,s,t}(a_1 \mid \mathcal{D}_t)$, $a_1 \in \mathcal{A}_{1,s}$,
where as in the main paper $\mathcal{D}_t$ comprises the data from
subjects enrolling prior to week $t$, and we suppress the damping
constant for brevity.  Application of the martingale central limit
theorem requires that the $\wone_i(a_1)$, which depend only on
$\mathcal{D}_t$ and not on data on subject $i$, are chosen so that (i)
the estimating equation (\ref{eq:Mest1}) remains conditionally (on
$\mathcal{D}_t$) unbiased,
$E[ M\{ A_{1i}, R_{1i}, Y_{1i}, a_1; \phi_{1,s}(a_1)\} \mid
\mathcal{D}_t ] = 0$; and (ii) the variance is stabilized,
$$E[ M^2\{ A_{1i}, R_{1i}, Y_{1i}, a_1; \phi_{1,s}(a_1)\} \mid
\mathcal{D}_t ] = \sigma^2 > 0,$$ a constant not depending on $t$.
With $\wone_i(a_1)$ so chosen, it then follows that the large sample
distribution of $\hatphi_{1,s}(a_1)$ can approximated as described
below.

  To show (i), note by the consistency assumption that a summand in
  (\ref{eq:Mest1}) can be written equivalently as
  $$ \wone_i(a_1)  I(A_{1i}=a_1)[ I\{ R_{1i}^*(a_1)=1, Y_{1i}^*(a_1)=1\}
  - \phi_{1,s}(a_1)].$$
  Note then that
   \begin{align}
 E &\left( \wone_i(a_1)  I(A_{1i}=a_1)[ I\{ R_{1i}^*(a_1)=1,
     Y_{1i}^*(a_1)=1\}  - \phi_{1,s}(a_1)] \left| \vphantom{E [ \wone_i(a_1)]} \right.   \mathcal{D}_t \right)  \nonumber \\
 &= E\left\{ E \left( \wone_i(a_1)  I(A_{1i}=a_1)[ I\{ R_{1i}^*(a_1)=1,
     Y_{1i}^*(a_1)=1\}  - \phi_{1,s}(a_1)] \left| \vphantom{E [ \wone_i(a_1)]} \right.   \mathcal{D}_t, \mathcal{W}_i^* \right)
 \left| \vphantom{E [ \wone_i(a_1)]} \right. \mathcal{D}_t \right\} \nonumber \\
 &=  E \left( \wone_i(a_1)  [ I\{ R_{1i}^*(a_1)=1,
     Y_{1i}^*(a_1)=1\}  - \phi_{1,s}(a_1)] \, E \{  I(A_{1i}=a_1) \mid \mathcal{D}_t, \mathcal{W}_i^* \}
 \left| \vphantom{E [ \wone_i(a_1)]} \right. \mathcal{D}_t \right) \nonumber \\
&= \wone_i(a_1) \pi_{1,s,t}(a_1 \mid \mathcal{D}_t) \, E \left(  [ I\{R_{1i}^*(a_1)=1,
    Y_{1i}^*(a_1)=1\}  - \phi_{1,s}(a_1)]  \mid\mathcal{D}_t \right) \label{eq:w1} \\
&= \wone_i(a_1) \pi_{1,s,t}(a_1 \mid \mathcal{D}_t) \,
E (  [ I\{R_{1i}^*(a_1)=1, Y_{1i}^*(a_1)=1\}  - \phi_{1,s}(a_1) ] = 0.\label{eq:zero}
   \end{align}
   Here, (\ref{eq:w1}) follows by showing that
   $E \{ I(A_{1i}=a_1) \mid \mathcal{D}_t, \mathcal{W}_i^* \} =
   \pi_{1,s,t}(a_1 \mid \mathcal{D}_t)$ by an argument conditional on
   $\mathcal{D}_t$ similar to that in of \citet[Section
   6.4.3]{tsiatis2020dynamic}, noting that $i$ is of subtype $s$.  The
   first equality in (\ref{eq:zero}) follows because
   $\mathcal{D}_t \independent \mathcal{W}_i^*$, where
   ``$\independent$'' denotes ``independent of'' (i.e., the data from
   past subjects are independent of the potential outcomes for subject
   $i$); and the second equality follows because in the ISPY2.2 SMART
   randomization at the first stage guarantees that
   $A_{1i} \independent \mathcal{W}_i^*$ given that $i$ is of subtype
   $s$ (as in the sequential randomization assumption).  Thus, using
   the consistency assumption,
   \begin{align*}
P\{ R_{1i}^*(a_1)=1,& Y_{1i}^*(a_1)=1 \} = P\{ R_{1i}^*(a_1)=1,
                      Y_{1i}^*(a_1)=1 \mid A_{1i} = a_1 \} \\
     &= P(R_{1i} =1, Y_{1i} = 1 \mid  A_{1i} = 1) = P(R_1=1, Y_1=1
       \mid  S=s, A_1=a_1) = \phi_{1,s}(a_1),
     \end{align*}
     given that $i$ is from subtype $s$.  Thus (i) holds for any
     $\wone_i(a_1)$ depending only on $\mathcal{D}_t$.  To determine
     $\wone_i(a_1)$ depending on $\mathcal{D}_t$ such that (ii) is
     satisfied, note that, similarly to the above, and using
     $\{I(A_{1i}=a_1)\}^2 = I(A_{1i}=a_1)$,
  \begin{align}
E &\left( \{ \wone_i(a_1)\}^2  I(A_{1i}=a_1)[ I\{ R_{1i}^*(a_1)=1,
    Y_{1i}^*(a_1)=1\}   - \phi_{1,s}(a_1)]^2\left| \vphantom{\{ \wone_i(a_1)\}^2} \right. \mathcal{D}_t \right)  \nonumber \\
&= E\left(  \{ \wone_i(a_1) \}^2  [ I\{ R_{1i}^*(a_1)=1,
    Y_{1i}^*(a_1)=\}  - \phi_{1,s}(a_1)]^2 \, E \{  I(A_{1i}=a_1) \mid  \mathcal{D}_t, \mathcal{W}_i^* \}
\left| \vphantom{\{ \wone_i(a_1)\}^2} \right. \mathcal{D}_t \right) \nonumber \\
   &=  \{ \wone_i(a_1) \}^2 \pi_{1,s,t}(a_1 \mid \mathcal{D}_t) \,
     E \left(  [ I\{(R_{1i}^*(a_1)=1,  Y_{1i}^*(a_1)=1\}  -
     \phi_{1,s}(a_1)]^2  \mid \mathcal{D}_t \right) \nonumber \\
 &=  \{ \wone_i(a_1) \}^2\pi_{1,s,t}(a_1 \mid \mathcal{D}_t) \,
 E \left(  [  I\{R_{1i}^*(a_1)=1, Y_{1i}^*(a_1)1\}  - \phi_{1,s}(a_1)]^2 \right), \label{eq:const}
   \end{align}
   where (\ref{eq:const}) follows because
   $\mathcal{D}_t \independent \mathcal{W}_i^*$.  Clearly, because the
   $\mathcal{W}_i^*$ are i.i.d.,
   $$E \left( [ I\{ R_{1i}^*(a_1)=1, Y_{1i}^*(a_1)=1\} -
     \phi_{1,s}(a_1)]^2\right)$$ is the same constant for any $i$ (and
   thus $t$).  Thus, the entire expression in (\ref{eq:const}) will be a
   constant not depending on $t$ if we take
\begin{equation}
\wone_i(a_1)  = 1 / \sqrt{ \pi_{1,s,t}(a_1 \mid \mathcal{D}_t) }.
  \label{eq:wonea1}
  \end{equation}

An entirely similar argument can be made for estimation of
$\phi_{2,s}(a_1)$.  Namely, letting
$$\hatalpha^w_{6,s}(a_1) =  n^{-1}_s\sumins \wone_i(a_1)  I(A_{1i}=a_1, R_{1i}=0),$$
the estimator
$\hatphi_{2,s}(a_1) = \hatalpha^w_{6,s}(a_1)/\hatalpha^w_{2,s}(a_1)$
for $\phi_{2,s}(a_1)$ in (\ref{eq:phi2}) solves in $\phi_{2,s}(a_1)$
an M-estimating equation of the form
\begin{equation}
\sumins \wone_i(a_1)  I(A_{1i}=a_1)\{ I(R_{1i}=0 -
\phi_{2,s}(a_1)\} = \sumins M\{ A_{1i}, R_{1i}, a_1; \phi_{2,s}(a_1)\} = 0.
\label{eq:Mest2}
\end{equation}
Arguments analogous to those above show that $\wone_i(a_1)$ should be
chosen as in (\ref{eq:wonea1}).

Similar arguments are possible for estimation of $\phi_{3,s}(a_1,a_2)$
and $\phi_{4,s}(a_1,a_2)$ in (\ref{eq:phi3}) and (\ref{eq:phi4}).
For a set of weights $\wtwo_i(a_1,a_2)$, $i = 1,\ldots,n_s$, define
$$\hatalpha^w_{3,s}(a_1,a_2) =  n^{-1}_s\sumins \wtwo_i(a_1,a_2)
I(A_{1i}=a_1, R_{1i}=0, A_{2i}=a_2, R_{2i}= 1, Y_{2i}=1),$$
$$\hatalpha^w_{4,s}(a_1,a_2) = n^{-1}_s \sumins \wtwo_i(a_1,a_2) I(A_{1i}=a_1, R_{1i}=0, A_{2i}=a_2)$$
$$\hatalpha^w_{5,s}(a_1,a_2) =  n^{-1}_s\sumins \wtwo_i(a_1,a_2)
I(A_{1i}=a_1, R_{1i}=0, A_{2i}=a_2, R_{2i}= 0, Y_{3i}=1).$$ Then as
above, estimators
$\hatphi_{3,s}(a_1,a_2) =
\hatalpha^w_{3,s}(a_1,a_2)/\hatalpha^w_{4,s}(a_1,a_2)$, 
$\hatphi_{4,s}(a_1,a_2) = \hatalpha^w_{5,s}(a_1,a_2)/
\hatalpha^w_{4,s}(a_1,a_2)$ solve M-estimating equations
\begin{equation}
\sumins \wtwo_i(a_1,a_2)  I(A_{1i}=a_1, R_{1i}=0, A_{2i}=a_2)\{ I(R_{2i}=1, Y_{2i}=1) -
\phi_{3,s}(a_1,a_2)\} = 0.
\label{eq:Mest3}
\end{equation}
\begin{equation}
\sumins \wtwo_i(a_1,a_2)  I(A_{1i}=a_1, R_{1i}=0, A_{2i}=a_2)\{ I(R_{2i}=0, Y_{3i}=1) -
\phi_{4,s}(a_1,a_2)\} = 0.
\label{eq:Mest4}
\end{equation}
As above, assume that subject $i$ enrolled during the interval
$[u, u+1)$ and then reached stage 2 during the interval $[t, t+1)$,
where $u < t$, so was assigned stage 1 treatment according to
randomization probabilities $\pi_{1,s,u}(a_1 \mid \mathcal{D}_u)$,
$a_1 \in \mathcal{A}_{1,s}$, and stage 2 treatment according to
randomization probabilities
$\pi_{2,s,t}(a_2 \mid a_1, \mathcal{D}_t)$,
$a_2 \in \mathcal{A}_{2,s}$.  Note that
$\mathcal{D}_u \subset \mathcal{D}_t$. Application of the martingale
central limit theorem now requires that the $\wtwo_i(a_1, a_2)$, which
depend only on $\mathcal{D}_t$ and not data on subject $i$, are chosen
so that (i) the estimating equations (\ref{eq:Mest3}) and
(\ref{eq:Mest4}) are conditionally (on $\mathcal{D}_t$) unbiased, and
(ii) the variance is stabilized.  It then follows that with
$\wtwo_i(a_1,a_2)$ so chosen, the estimators $\hatphi_{3,s}(a_1, a_2)$
and $\hatphi_{4,s}(a_1, a_2)$ are consistent for $\phi_{3,s}(a_1,a_2)$
and $\phi_{4,s}(a_1,a_2)$ with the large sample distributions that can
be established via the martingale central limit theorem.

We demonstrate (i) and determine $\wtwo_i(a_1,a_2)$ depending on
$\mathcal{D}_t$ such that (ii) is satisfied for (\ref{eq:Mest3}); the
argument for (\ref{eq:Mest4}) is entirely similar and leads to the
same choice of $\wtwo_i(a_1,a_2)$.  First, using the consistency
assumption, a summand in (\ref{eq:Mest3}) can be written as
$$\wtwo_i(a_1,a_2)  I(A_{1i}=a_1, A_{2i}=a_2) I\{R_{1i}^*(a_1)=0\}
[ I\{R_{2i}^*(a_1,a_2)=1, Y_{2i}^*(a_1,a_2)=1 \} -\phi_{3,s}(a_1,a_2)].$$
Then (i) follows because
  \begin{align}
E&\left( \wtwo_i(a_1,a_2)  I(A_{1i}=a_1, A_{2i}=a_2) I\{R_{1i}^*(a_1)=0\}
[ I\{ R_{2i}^*(a_1,a_2)=1, Y_{2i}^*(a_1,a_2)=1\} -\phi_{3,s}(a_1,a_2)] \left| \vphantom{\wtwo_i(a_1,a_2)}
                 \right. \mathcal{D}_t\right) \nonumber \\
&= E\left\{E\left( \wtwo_i(a_1,a_2)  I(A_{1i}=a_1, A_{2i}=a_2)  I\{R_{1i}^*(a_1)=0\}
                                                             \right.\right.\nonumber \\
&\hspace{0.2in}\times \left.\left. [ I\{ R_{2i}^*(a_1,a_2)=1, Y_{2i}^*(a_1,a_2)=1\} -\phi_{3,s}(a_1,a_2)] \left| \vphantom{\wtwo_i(a_1,a_2)}
                 \right. \mathcal{D}_t, \mathcal{W}^*_i\right) \left| \vphantom{\wtwo_i(a_1,a_2)}
                 \right. \mathcal{D}_t \right\}\nonumber \\
&=  E\left( \wtwo_i(a_1,a_2)  I\{R_{1i}^*(a_1)=0\} [
                                                              I\{R_{2i}^*(a_1,a_2)=1, Y_{2i}^*(a_1,a_2)= 1] -\phi_{3,s}(a_1,a_2)] \right.\nonumber \\
&\hspace{0.2in}\times \left. E\{ I(A_{1i}=a_1, A_{2i}=a_2) \mid \mathcal{D}_t, \mathcal{W}^*_i\} \left| \vphantom{\wtwo_i(a_1,a_2)}
                 \right. \mathcal{D}_t \right) \nonumber \\
&=  E\left( \wtwo_i(a_1,a_2) \pi_{1,s,u}(a_1 \mid \mathcal{D}_u) \pi_{2,s,t}(a_2 \mid a_1, \mathcal{D}_t)
                                                              I\{R_{1i}^*(a_1)=0\} \right. \nonumber \\
 &\hspace{0.2in}\times \left.   [ I\{ R_{2i}^*(a_1,a_2)=1, Y_{2i}^*(a_1,a_2)=1\} -\phi_{3,s}(a_1,a_2)]
 \left| \vphantom{\wtwo_i(a_1,a_2)}  \right. \mathcal{D}_t \right) \label{eq:phi31}\\
 &=  \wtwo_i(a_1,a_2) \pi_{1,s,u}(a_1 \mid \mathcal{D}_u) \pi_{2,s,t}(a_2 \mid a_1, \mathcal{D}_t)
   E\big( I\{R_{1i}^*(a_1)=0\} \nonumber \\
  &\hspace{0.2in}\times  [ I\{ R_{2i}^*(a_1,a_2)=1, Y_{2i}^*(a_1,a_2)=1\} -\phi_{3,s}(a_1,a_2)] \big) =0.
 \label{eq:phi32} 
 \end{align}
 Here, as above the equality in (\ref{eq:phi31}) follows because
 $E\{ I(A_{1i}=a_1, A_{2i}=a_2) \mid \mathcal{D}_t, \mathcal{W}^*_i\}
 = \pi_{1,s,u}(a_1 \mid \mathcal{D}_u) \pi_{2,s,t}(a_2 \mid a_1,
 \mathcal{D}_t)$ by an argument conditional on $ \mathcal{D}_t$
 similar to that in of \citet[Section 6.4.3]{tsiatis2020dynamic}.  The first equality in
 (\ref{eq:phi32}) holds because
 $\mathcal{D}_t \independent \mathcal{W}_i^*$ as above.  The second
 equality in (\ref{eq:phi32}) holds because, using
 $E [ I\{R_{1i}^*(a_1)=0\} ]= P\{ R_{1i}^*(a_1)=0\}$, the consistency
 assumption, and the fact that randomization guarantees
 $A_{1i} \independent \mathcal{W}_i^*$ and
 $A_{2i} \independent \mathcal{W}_i^* \mid R_{1i}=0,A_{1i}$ given $i$ is
 also of subtype $s$ (as in the sequential randomization assumption)
 \begin{align*}
   P&\{ R_{1i}^*(a_1)=0, R_{2i}^*(a_1,a_2)=1, Y_{2i}^*(a_1,a_2)=1 \}\\
&=   P\{ R_{1i}^*(a_1)=0, R_{2i}^*(a_1,a_2)=1, Y_{2i}^*(a_1,a_2)=1 \mid A_{1i}=a_1\}\\
&= P\{ R_{2i}^*(a_1,a_2)=1, Y_{2i}^*(a_1,a_2)=1 \mid A_{1i}=a_1, R_{1i}^*(a_1)=0 \}\,
     P\{ R_{1i}^*(a_1)=0 \mid A_{1i}=a_1\}\\
&= P\{ R_{2i}^*(a_1,a_2)=1, Y_{2i}^*(a_1,a_2)=1 \mid A_{1i}=a_1, R_{1i}^*(a_1)=0\}\,\,
                                                                                                    P\{ R_{1i}^*(a_1)=0\} \\
&= P\{ R_{2i}^*(a_1,a_2)=1, Y_{2i}^*(a_1,a_2)=1 \mid A_{1i}=a_1, R_{1i}=0\}\,\, P\{ R_{1i}^*(a_1)=0\} \\
   &= P\{ R_{2i}^*(a_1,a_2)=1, Y_{2i}^*(a_1,a_2)=1 \mid A_{1i}=a_1, R_{1i}=0, A_{2i}=a_2\}\,\,
                                                                                                    P\{ R_{1i}^*(a_1)=0\} \\
&= P( R_{2i}=1, Y_{2i} = 1 \mid A_{1i}=a_1, R_{1i}=0, A_{2i}=a_2) P\{ R_{1i}^*(a_1)=0\}\\
               &= P(R_2=1, Y_2=1 \mid S=s, A_1=a_1, R_1=0, A_2=a_2) \,P\{ R_{1i}^*(a_1)=0\} = \phi_{3,s}(a_1,a_2) P\{ R_{1i}^*(a_1)=0\}.
 \end{align*}

 Finally, to determine $\wtwo_i(a_1,a_2)$ depending on $\mathcal{D}_t$
 such that (ii) is satisfied, by similar arguments,
\begin{align}
E&\left( \{\wtwo_i(a_1,a_2)\}^2  I(A_{1i}=a_1, A_{2i}=a_2)
   I\{R_{1i}^*(a_1)=0\} \right. \nonumber \\
&\hspace{0.2in}\times \left.[ I\{ R_{2i}^*(a_1,a_2)=1, Y_{2i}^*(a_1,a_2)=1\} -\phi_{3,s}(a_1,a_2)]^2
\left| \vphantom{\wtwo_i(a_1,a_2)} \right. \mathcal{D}_t\right) \nonumber \\
&=E\left\{ E \left( \{\wtwo_i(a_1,a_2)\}^2  I(A_{1i}=a_1, A_{2i}=a_2)    I\{R_{1i}^*(a_1)=0\}   \right. \right.\nonumber \\
&\hspace{0.2in}\times \left. \left.[ I\{ R_{2i}^*(a_1,a_2)=1, Y_{2i}^*(a_1,a_2)=1\} -\phi_{3,s}(a_1,a_2)]^2
   \left| \vphantom{\wtwo_i(a_1,a_2)} \right. \mathcal{D}_t,
     \mathcal{W}^*\right)\left| \vphantom{\wtwo_i(a_1,a_2)}  \right. \mathcal{D}_t
   \right\}\nonumber  \\
&= E\left( \{\wtwo_i(a_1,a_2)\}^2  I\{R_{1i}^*(a_1)=0\} [ I\{ R_{2i}^*(a_1,a_2)=1, Y_{2i}^*(a_1,a_2)=1\}
                           -\phi_{3,s}(a_1,a_2)]^2\right. \nonumber \\
 &\hspace{0.2in}\times \left. E\{ I(A_{1i}=a_1, A_{2i}=a_2) \mid \mathcal{D}_t,  \mathcal{W}^*_i\} \left|
                          \vphantom{\wtwo_i(a_1,a_2)}    \right. \mathcal{D}_t \right)\nonumber \\
&= \{\wtwo_i(a_1,a_2)\}^2 \pi_{1,s,u}(a_1 \mid \mathcal{D}_u) \pi_{2,s,t}(a_2 \mid a_1,
 \mathcal{D}_t) \nonumber \\ &\hspace{0.2in}\times E\left( I\{R_{1i}^*(a_1)=0\}[ I\{ R_{2i}^*(a_1,a_2)=1, Y_{2i}^*(a_1,a_2)=1\} -\phi_{3,s}(a_1,a_2)]^2
\left|  \vphantom{\wtwo_i(a_1,a_2)}    \right. \mathcal{D}_t    \right)\nonumber\\
&= \{\wtwo_i(a_1,a_2)\}^2 \pi_{1,s,u}(a_1 \mid \mathcal{D}_u) \pi_{2,s,t}(a_2 \mid a_1,
 \mathcal{D}_t) \nonumber \\ &\hspace{0.2in}\times E\left( I\{R_{1i}^*(a_1)=0\}[ I\{ R_{2i}^*(a_1,a_2)=1, Y_{2i}^*(a_1,a_2)=1\} -\phi_{3,s}(a_1,a_2)]^2\right).
\label{eq:w2.2}
 \end{align} 
Because $\mathcal{W}_i$, $i=1,\ldots,n_s$, are i.i.d., the expectation
(\ref{eq:w2.2}) is a constant not depending on $t$.  Thus, the entire
expression will be constant and not depending on $t$ if we take
\begin{equation}
\wtwo_i(a_1)  = 1 / \sqrt{ \pi_{1,s,u}(a_1 \mid \mathcal{D}_u) \pi_{2,s,t}(a_2 \mid a_1,
 \mathcal{D}_t) }.
  \label{eq:wtwoa1a2}
  \end{equation}
 
  Based on these results, the proposed weighted estimator for
  $\mu_s(a_1,a_2)$ is given by
  \begin{equation}
    \label{eq:muwtsamp}
    \begin{aligned}
      \hatmu_{s,wtsamp}(a_1, a_2) &= \hatphi_{1,s}(a_1) +
      \hatphi_{2,s}(a_1) \{ \hatphi_{3,s}(a_1,a_2) + \hatphi_{4,s}(a_1, a_2)\} \\
      &= \frac{\hatalpha^w_{1,s}(a_1)}{\hatalpha^w_{2,s}(a_1)} +
      \frac{\hatalpha^w_{6,s}(a_1)}{\hatalpha^w_{2,s}(a_1)} \left\{
        \frac{\hatalpha^w_{3,s}(a_1,a_2)+\hatalpha^w_{5,s}(a_1,a_2)}{\hatalpha^w_{4,s}(a_1,a_2)}\right\}.
\end{aligned}
\end{equation}

Similar to \citet{ZhangMest2021} and \citet{NorwoodRAR}, from the above, under regularity
conditions and by the martingale central limit theorem, suitably scaled versions of
$n_s^{1/2} \{ \hatphi_{1,s}(a_1) -\phi_{1.s}(a_1) \}$,
$n_s^{1/2} \{ \hatphi_{2,s}(a_1) -\phi_{3.s}(a_1) \}$ and $n_s^{1/2} [
\{ \hatphi_{3,s}(a_1,a_2) + \hatphi_{4,s}(a_1, a_2)\} - \{
\phi_{3,s}(a_1, a_2) + \phi_{4,s}(a_1, a_2) \}$ converge in
distribution to standard normal random variables.  We can write
\begin{equation}
  \label{eq:wtsampsum}
\begin{aligned}
n^{1/2}_s&\left\{ \hatmu_{s,wtsamp}(a_1,a_2) - \mu_s(a_1, a_2)\right\} =
n^{1/2}_s \{ \hatphi_{1,s}(a_1) -\phi_{1.s}(a_1) \} +
\{\phi_{3,s}(a_1, a_2) \\ &+ \phi_{4,s}(a_1, a_2) \}
           n^{1/2}_s \{ \hatphi_{2,s}(a_1) -\phi_{2,s}(a_1) \} \\
  &+ \phi_{2,s}(a_1)  n^{1/2}_s [ \{ \hatphi_{3,s}(a_1,a_2) + \hatphi_{4,s}(a_1,
           a_2)\} - \{\phi_{3,s}(a_1, a_2) + \phi_{4,s}(a_1, a_2) \}]
           + o_p(1).
         \end{aligned}
         \end{equation}
         Assuming that the scaling factors converge in probability and
         are well-behaved,
         $n_s^{1/2} \{ \hatphi_{1,s}(a_1) -\phi_{1.s}(a_1) \}$,
         $n_s^{1/2} \{ \hatphi_{2,s}(a_1) -\phi_{3.s}(a_1) \}$ and
         $n_s^{1/2} [ \{ \hatphi_{3,s}(a_1,a_2) + \hatphi_{4,s}(a_1,
         a_2)\} - \{ \phi_{3,s}(a_1, a_2) + \phi_{4,s}(a_1, a_2) \}]$
         each themselves converge in distribution to normal random
         variables, and thus so does (\ref{eq:wtsampsum}). Via tedious
         arguments, it can be shown that $n^{1/2}_s\left\{
           \hatmu_{s,wtsamp}(a_1,a_2) - \mu_s(a_1, a_2)\right\}$
         converges in distribution to a mean-zero normal random
         variable with variance $\Sigma^w >0$, where $\Sigma^w$ is the
         limit in probability of $n_s^{-1} \sumins
         \mathcal{I}_i^{w\,2}$,  and
\begin{align}
\mathcal{I}^w_i &= \left\{ \frac{1}{\alpha_{2,s}(a_1)}\right\} \left( I(A_{1i}=a_1, R_{1i}=1, Y_{1i}=1) - \alpha_{1,s}(a_1) -
  \mu_s(a_1, a_2)\{ I(A_{1i}=a_1) - \alpha_{2,s}(a_1)\} \vphantom{\left(\frac{\alpha_{3,s}(a_1,a_2)+\alpha_{5,s}(a_1,a_2)}{\alpha_{4,s}(a_1,a_2)}\right)}\right. \nonumber \\
  &\hspace{0.05in}+ \left\{ \frac{\alpha_{3,s}(a_1,a_2)+\alpha_{5,s}(a_1,a_2)}{\alpha_{4,s}(a_1,a_2)}\right\} \left[
    -\left\{\frac{\alpha_{6,s}(a_1)}{\alpha_{4,s}(a_1,a_2)}\right\}\left\{ I(A_{1i}=a_1,R_{1i}=0,
    A_{2i}=a_2) - \alpha_{4,s}(a_1,a_2)\right\} \right.\nonumber \\
  &\hspace{0.15in}+\left.
    \left\{I(A_{1i}=a_1,R_{1i}=0) - \alpha_{6,s}(a_1)\right\}
    \vphantom{ \frac{\alpha_{6,s}(a_1)}{\alpha_{4,s}(a_1,a_2)}}\right] \nonumber \\
  &\hspace{0.05in}+ \left.\left\{ \frac{\alpha_{6,s}(a_1)}{\alpha_{4,s}(a_1,a_2)}\right\}
    \left[ \vphantom{ \{\alpha_{3,s}(a_1,a_2)+\alpha_{5,s}(a_1,a_2\} }
      I(A_{1i}=a_1, R_{1i}=0, A_{2i}=a_2, R_{2i}=1,Y_{2i}=1) \right. \right.\nonumber  \\
  &\hspace{0.15in}+
      I(A_{1i}=a_1, R_{1i}=0, A_{2i}=a_2, R_{2i}=0, Y_{3i}=1)  \nonumber \\
&\hspace*{0.85in} \left. \left. - \{\alpha_{3,s}(a_1,a_2)+\alpha_{5,s}(a_1,a_2\}
    \vphantom{ \left[  I(A_{1i}=a_1, R_{1i}=0, A_{2i}=a_2,
        R_{2i}=1,Y_{2i}=1) + I(A_{1i}=a_1, R_{1i}=0, A_{2i}=a_2,
        R_{2i}=0, Y_{3i}=1)\right]}  \right] \vphantom{ \left\{
      \frac{\alpha_{6,s}(a_1)}{\alpha_{4,s}(a_1,a_2)}\right\} }    \right).
  \label{eq:ifw}
\end{align}
For practical use, letting $\widehat{\mathcal{I}}^w_i$ denote
(\ref{eq:ifw}) with
$\hatalpha^w_{1,s}(a_1), \hatalpha^w_{2,s}(a_1), \hatalpha^w_{3,s}(a_1,a_2),
\hatalpha^w_{4,s}(a_1,a_2)$, $\hatalpha^w_{5,s}(a_1,a_2),
\hatalpha^w_{6,s}(a_1)$ and $\hatmu_{s,wtsamp}(a_1, a_2)$ substituted for
$\alpha_{1,s}(a_1), \alpha_{2,s}(a_1), \alpha_{3,s}(a_1, a_2), \alpha_{4,s}(a_1,
a_2)$, $\alpha_{5,s}(a_1, a_2), \alpha_{6,s}(a_1)$ and $\mu_s(a_1, a_2)$,
estimate $\Sigma^w$ by $n^{-1}_s \sumins \widehat{\mathcal{I}}_i^{w\,2}$.  

\setcounter{equation}{0}
\renewcommand{\theequation}{C.\arabic{equation}}

\section{C. Derivation of (15)}\label{s:valuederivation}

We present the derivation of the expression in (\ref{eq:simvalue}) of
the main paper for the true value of regime $\{a_1, a_2\}$, namely,
 \begin{equation}
   \label{eq:simvalue15}
   \mu_s(a_1, a_2) = p_1(a_1) + 
p_2(a_1,a_2)\{1-p_1(a_1)\} \lambda_{spec} +
p_3(a_1,a_2) \{1- p_2(a_1,a_2)\}\{1-p_1(a_1)\} \lambda_{spec}^2.
%   \begin{aligned}
% \mu_s(a_1, a_2) = p_1(a_1) &\lambda_{sens} + 
% p_1(a_1) \lambda_{sens} (1- \lambda_{sens}) + 
% p_2(a_1,a_2) \{1-p_1(a_1)\} \lambda_{sens} \lambda_{spec} +
% p_1(a_1) (1-\lambda_{sens})^2 \\ &+ p_2(a_1,a_2) \{1-p_1(a_1)\} (1-\lambda_{sens})
% \lambda_{spec} +    p_3(a_1,a_2) \{1- p_2(a_1,a_2)\}\{1-p_1(a_1)\} \lambda_{spec}^2,
%   \end{aligned}
\end{equation}
under several assumptions we now state.  Throughout this section, we
suppress the event $S=s$ for brevity.  

As defined in the main paper, the durability assumption states that,
if a subject achieves pCR after stage 1, $Y_1 = 1$, but $R_1 = 0$, so
that the subject proceeds to stage 2, then $Y_2=1$; similarly, if a
subject achieves pCR after stage 1 or 2, so that $Y_1=1, Y_2=1$ or
$Y_1=0, Y_2=1$, but $R_1=R_2=0$, so that the subject proceeds to stage
3, then $Y_3=1$.  Formally, the durability assumption thus implies
that
\begin{equation}
  \label{eq:durability}
  \begin{aligned}
&P(Y_2 = 1 \mid A_1=a_1, Y_1=1, R_1=0, A_2=a_2) = 1\\
P(Y_3=1 \mid  &A_1=a_1, Y_1=0, R_1=0, A_2=a_2, Y_2=1, R_2 = 0) = 1 \\
PY_3=1 \mid &A_1=a_1, Y_1=1, R_1=0, A_2=a_2, Y_2=1, R_2 = 0) = 1,
\end{aligned}
\end{equation}
and that the event $(Y_3=1, Y_2=0, Y_1 = 1)$ and similar events occur
with probability zero.  The sensitivity and sensitivity of the preRCB algorithm
as in the main paper are assumed to be independent of treatment
and prior pCR status; that is,
\begin{equation}
  \label{eq:sens}
  \begin{aligned}
  \lambda_{sens} &= P(R_1=1 \mid A_1=a_1, Y_1=1) \\
  &= P(R_2=1 \mid A_1=a_1,Y_1=0, R_1=0, A_2=a_2,Y_2 =1) \\
  &= P(R_2=1 \mid A_1=a_1,Y_1=1, R_1=0, A_2=a_2,Y_2 =1) = 0.53
\end{aligned}
\end{equation}
\begin{equation}
  \label{eq:spec}
  \begin{aligned}
    \lambda_{spec} &= P(R_1=0 \mid A_1=a_1, Y_1=0) \\
    &= P(R_2 = 0 \mid A_1=a_1, Y_1=0, R_1=0, A_2=a_2, Y_2=0) \\
    &= P(R_2 = 0 \mid A_1=a_1, Y_1=1, R_1=0, A_2=a_2, Y_2=0)  = 0.90
  \end{aligned}
\end{equation}
for all $(a_1, a_2) \in \mathcal{A}_{1,s} \times \mathcal{A}_{2,s}$.
Regarding the true pCR rates $p_1(a_1)$, $p_2(a_1,a_2)$,
$p_3(a_1,a_2)$ defined in the main paper, we make the following
surrogacy assumptions, which are implicit in the generative data
process used for the simulations; namely,
\begin{equation}
  \label{eq:surrogacy}
  \begin{aligned}
    p_2(a_1,a_2) & = P(Y_2 = 1 \mid A_1 = a_1, Y_1=0, A_2=a_2) = P(Y_2
    = 1 \mid A_1 = a_1, Y_1=0, R_1=0,  A_2=a_2)\\
    p_3(a_1,a_2) &= P(Y_3 = 1 \mid A_1=a_1, Y_1=0, A_2=a_2, Y_2=0)\\
    &= P(Y_3 = 1 \mid A_1=a_1, Y_1=0, R_1=0, A_2=a_2 Y_2=0, R_2=0), 
\end{aligned}
\end{equation}
for $(a_1,a_2) \in \mathcal{A}_{1,s} \times\mathcal{A}_{2,s}$; that
is, for any stage 1 and 2 treatments and given that a patient has not
yet achieved pCR, the result of preRCB testing has no bearing on a
patient's current true pCR status.  Finally, because simple and
adaptive randomization to stage 2 treatment for a patient for whom
$R_1 = 0$ depends only on a patient's stage 1 treatment and, in the
latter case, data from previous participants, and not on the patient's
true pCR status $Y_1$,
\begin{equation}
P(A_2 = a_2 \mid A_1 = a_1, Y_1=y_1, R_1=0) = P(A_2 = a_2 \mid A_1 =
a_1, R_1=0), \hspace*{0.15in} y_1 = 0, 1.
\label{eq:random}
  \end{equation}

With the definitions in (\ref{eq:thetagamma1.supp})-(\ref{eq:gamma3.supp}), from
(\ref{eq:value.supp}), as in (6) of  the main paper, the true value of
regime $\{a_1, a_2\}$ is given by
\begin{equation}
   \label{eq:valueagain} 
\begin{aligned}
 \mu_s(a_1,a_2) &= \theta_1(a_1) \gamma_1(a_1) +
 \{1-\theta_1(a_1)\} \theta_2(a_1,a_2) \gamma_2(a_1,a_2) \\
&+ \{1-\theta_1(a_1)\}\{1-\theta_2(a_1,a_2)\}
\gamma_{3,s}(a_1,a_2). 
\end{aligned}
\end{equation}
To demonstrate (\ref{eq:simvalue15}) under the foregoing assumptions,
we reexpress each term in (\ref{eq:valueagain}) in terms of the above quantities.

First, it is straightforward, using (\ref{eq:sens}) and the definition
of $p_1(a_1)$ that the first term on the right hand side of
(\ref{eq:valueagain}) can be written as
\begin{equation}
\label{eq:piece1}
  \begin{aligned}
\theta_1(a_1) \gamma_1(a_1)&= P(Y_1 = 1, R_1=1 \mid A_1=a_1) \\ &=
P(R_1=1 \mid A_1=a_1, Y_1 = 1)  P((Y_1 = 1 \mid A_1=a_1) =
\lambda_{sens} \, p_1(a_1), 
\end{aligned}
\end{equation}
and $\{ 1 - \theta_{1,s}(a_1) \} = P(R_1 = 0 \mid
A_a=a_1)$.  To reexpress the second term in (\ref{eq:valueagain}), note that
\begin{equation}
  \label{eq:piece21}
\begin{aligned}
  &\theta_{2,s}(a_1, a_2) \gamma_{2,s}(a_1, a_2)\} = P(Y_2=1, R_2=1
  \mid A_1=a_1, R_1=0, A_2=a_2) \\
  &= P(Y_2=1, R_2=1, Y_1=0 \mid A_1=a_1, R_1=0, A_2=a_2) \\ &+
  P(Y_2=1, R_2=1, Y_1=1 \mid A_1=a_1, R_1=0, A_2=a_2) \\
  &= P(R_2=1 \mid A_1=a_1,Y_1=0, R_1=0, A_2=a_2,Y_2 =1) \, P(Y_2=1
  \mid A_1 = a_1, Y_1=0, R_1=0,  A_2=a_2) \\
  &\hspace*{0.3in}\times P(Y_1=0 \mid  A_1 = a_1,  R_1=0,  A_2=a_2)\\
  &\hspace*{0.1in}+ P(R_2=1 \mid A_1=a_1,Y_1=1, R_1=0, A_2=a_2,Y_2 =1) \, P(Y_2=1
  \mid A_1 = a_1, Y_1=1, R_1=0,  A_2=a_2) \\
  &\hspace*{0.3in}\times P(Y_1=1 \mid  A_1 = a_1,  R_1=0,  A_2=a_2)\\
  &=\lambda_{sens} \, p_2(a_1, a_2) P(Y_1=0 \mid  A_1 = a_1,
  R_1=0,  A_2=a_2) + \lambda_{sens} \, P(Y_1=1 \mid  A_1 = a_1,
  R_1=0,  A_2=a_2),
  \end{aligned}
\end{equation}
where the final equality follows from (\ref{eq:durability}) and
(\ref{eq:sens}).  Now it is straightforward that
\begin{equation}
  \label{eq:final1}
\begin{aligned}
  P&(Y_1=0 \mid  A_1 = a_1,  R_1=0,  A_2=a_2) \\
  &= \left\{ \frac{ P(A_2 = a_2 \mid A_1 = a_1, Y_1=0, R_1=0) }{P(A_2 = a_2 \mid A_1 =
      a_1, R_1=0)} \right\} \,\left\{ \frac{ P(R_1=0 \mid A_1=a_1,
  Y_1=0) P(Y_1=0 \mid A_1=a_1) }{P(R_1=0 \mid A_1=a_1) } \right\} \\
&= \frac{\lambda_{spec} \{ 1-p_1(a_1)\} }{ 1-\theta_{1,s}(a_1) }
\end{aligned}
\end{equation}
 using (\ref{eq:spec}) and (\ref{eq:random}).   By an entirely similar
 argument,
 \begin{equation}
 P(Y_1 = 1 \mid  A_1 = a_1,  R_1=0,  A_2=a_2) = \frac{
   (1-\lambda_{sens}) p_1 (a_1)} {1-\theta_{1,s}(a_1) }.
 \label{eq:final2}
\end{equation}
 Combining these results with (\ref{eq:piece21}) then yields
 \begin{equation}
   \{1-\theta_1(a_1)\} \theta_2(a_1,a_2) \gamma_2(a_1,a_2) 
   = p_2(a_1,a_2) \{1-p_1(a_1)\} \lambda_{sens} \lambda_{spec} +
   p_1(a_1) \lambda_{sens} (1- \lambda_{sens}).
   \label{eq:piece2}
 \end{equation}
 To reexpress the third term in (\ref{eq:valueagain}), note that
 \begin{equation}
   \label{eq:piece30}
   \begin{aligned}
\{1-&\theta_2(a_1,a_2)\} \gamma_{3,s}(a_1,a_2) = P(Y_3=1, R_2=0 \mid A_1=a_1, R_1=0, A_2=a_2) \\
&= P(Y_3=1, R_2=0, Y_1=0, Y_2=0 \mid A_1=a_1, R_1=0, A_2=a_2) \\
&\hspace{0.1in}+ P(Y_3=1, R_2= 0, Y_1=0, Y_2=1 \mid A_1=a_1, R_1=0, A_2=a_2) \\
&\hspace{0.1in}+ P(Y_3=1, R_2= 0, Y_1=1, Y_2=0 \mid A_1=a_1, R_1=0,
A_2=a_2)\\
&\hspace{0.1in}+ P(Y_3=1, R_2= 0, Y_1=1, Y_2=1 \mid A_1=a_1, R_1=0, A_2=a_2).
   \end{aligned}
   \end{equation}
   Because the event in the third term on the right hand side of
   (\ref{eq:piece30}) can never occur, this term is equal to zero.  We
   thus consider each of the remaining terms.  Using
   (\ref{eq:spec}), (\ref{eq:surrogacy}),  and (\ref{eq:final1}), the
   first term on the right hand side of (\ref{eq:piece30}) can be
   written as
   \begin{equation}
   \label{eq:piece31}
   \begin{aligned}
P&(Y_3=1,  R_2= 0, Y_1=0, Y_2=0 \mid A_1=a_1, R_1=0, A_2=a_2) \\
&= P(Y_3 = 1 \mid A_1=a_1, Y_1=0, R_1=0, A_2=a_2 Y_2=0) \\
 &\hspace{0.1in}\times P(R_2 = 0 \mid
 A_1=a_1, Y_1=0, R_1=0, A_2=a_2, Y_2=0)\\
 &\hspace{0.1in}\times P(Y_2 = 0 \mid A_1 = a_1, Y_1=0, R_1=0, A_2=a_2)
 P(Y_1=0 \mid  A_1 = a_1,  R_1=0,  A_2=a_2)\\
 &= p_3(a_1,a_2) \, \lambda_{spec} \, \{1-p_2(a_1,a_2)\} \frac{\lambda_{spec} \{ 1-p_1(a_1)\} }{ 1-\theta_{1,s}(a_1) }.
\end{aligned}
\end{equation}
Similarly, and also using (\ref{eq:durability}) and (\ref{eq:sens}),
the second and fourth terms can be written as
\begin{equation}
   \label{eq:piece32}
   \begin{aligned}
P&(Y_3=1,  R_2= 0, Y_1=0, Y_2=1 \mid A_1=a_1, R_1=0, A_2=a_2) \\
&= P(Y_3 = 1 \mid A_1=a_1, Y_1=0, R_1=0, A_2=a_2 Y_2 = 1) \\
&\hspace{0.1in}\times P(R_2 = 0 \mid
 A_1=a_1, Y_1=0, R_1=0, A_2=a_2, Y_2=1)\\
 &\hspace{0.1in}\times P(Y_2 = 1 \mid A_1 = a_1, Y_1=0, R_1=0, A_2=a_2) \,
 P(Y_1=0 \mid  A_1 = a_1,  R_1=0,  A_2=a_2)\\
 &= \, (1-\lambda_{sens}) p_2(a_1,a_2) \frac{\lambda_{spec} \{ 1-p_1(a_1)\} }{ 1-\theta_{1,s}(a_1) }
\end{aligned}
\end{equation}
and 
\begin{equation}
   \label{eq:piece34}
   \begin{aligned}
P&(Y_3=1,  R_2= 0, Y_1= 1, Y_2=1 \mid A_1=a_1, R_1=0, A_2=a_2) \\
 &= P(Y_3 = 1 \mid A_1=a_1, Y_1=1, R_1=0, A_2=a_2 Y_2=1) \\
 &\hspace{0.1in}\times P(R_2 = 0 \mid
 A_1=a_1, Y_1=1, R_1=0, A_2=a_2, Y_2=1)\\
 &\hspace{0.1in}\times P(Y_2 = 1 \mid A_1 = a_1, Y_1=1, R_1=0, A_2=a_2)
 P(Y_1=1 \mid  A_1 = a_1,  R_1=0,  A_2=a_2)\\
 &= \, (1-\lambda_{sens}) \frac{
   (1-\lambda_{sens}) p_1 (a_1)} {1-\theta_{1,s}(a_1) }.
\end{aligned}
\end{equation}
Substituting (\ref{eq:piece31}) - (\ref{eq:piece34}) in
(\ref{eq:piece30}), the third term in (\ref{eq:valueagain}) is given
by
\begin{equation}
\label{eq:piece3}
  \begin{aligned}
   &\{1-\theta_1(a_1)\} \{1-\theta_2(a_1,a_2) \}\gamma_3(a_1,a_2) \\
   &=  p_3(a_1,a_2) \{1- p_2(a_1,a_2)\}\{1-p_1(a_1)\} \lambda_{spec}^2
p_2(a_1,a_2) \{1-p_1(a_1)\} (1-\lambda_{sens}) \lambda_{spec} \\ &\hspace*{0.2in}+  
p_1(a_1) (1-\lambda_{sens})^2.
\end{aligned}
\end{equation}
Substituting (\ref{eq:piece1}), (\ref{eq:piece2}), and
(\ref{eq:piece3}) in (\ref{eq:valueagain}) and noting that
$$p_2(a_1,a_2) \{1-p_1(a_1)\} \lambda_{sens} \lambda_{spec} +
p_2(a_1,a_2) \{1-p_1(a_1)\} (1-\lambda_{sens}) \lambda_{spec} =
p_2(a_1,a_2) \{1-p_1(a_1)\}\lambda_{spec}$$ and
$p_1(a_1) \{\lambda_{sens} + \lambda_{sens} (1-\lambda_{sens}) +
(1-\lambda_{sens})^2\} = p_1(a_1)$ yields the expression for the true value in
(\ref{eq:simvalue15}), as desired.
 
\setcounter{equation}{0}
\renewcommand{\theequation}{D.\arabic{equation}}

\section{D. Additional simulation details and results}\label{s:additional}

\noindent
\textbf{D.1  Simulation details} \\*[0.1in]

\noindent{\em Sensitivity and specificity.}  Taking sensitivity and
specificity of the preRCB algorithm to be independent of treatment and
prior pCR status is characterized as
\begin{align*}
\lambda_{sens} &= P(R_1=1 \mid A_1=a_1, Y_1=1) = P(R_2=1 \mid
A_1=a_1,Y_1=0, R_1=0, A_2=a_2,Y_2 =1) \\ &= P(R_2=1 \mid A_1=a_1,Y_1=1,
  R_1=0, A_2=a_2,Y_2 =1) = 0.53
  \end{align*}
(sensitivity) and 
\begin{align*}\lambda_{spec}& = P(R_1=0 \mid A_1=a_1, Y_1=0) = P(R_2 = 0 \mid
A_1=a_1, Y_1=0, R_1=0, A_2=a_2, Y_2=0) \\ &= P(R_2 = 0 \mid A_1=a_1,
  Y_1=1, R_1=0, A_2=a_2, Y_2=0) = 0.90
\end{align*}
 (specificity) for all
$(a_1, a_2) \in \mathcal{A}_{1,s} \times \mathcal{A}_{2,s}$,
suppressing conditioning on $S=s$ for brevity.  These relationships
were enforced in all generative scenarios.

\vspace*{0.1in}

\noindent{\em Data generation details.}  Under all randomization schemes,
for each subject, at enrollment, $A_1 \in \mathcal{A}_{1,s}$ was
generated as Bernoulli using the current stage 1 randomization
probabilities.  At week 12 post enrollment, $Y_1$ was generated as
Bernoulli$\{p_1(a_1)\}$ for $A_1 = a_1$, and $R_1$ was generated as
Bernoulli$\{p_R(Y_1)\}$, where
$p_R(y) = \lambda_{sens} I(y=1) + (1-\lambda_{spec}) I(y=0)$.  If
$R_1=1$, $Y_1$ was recorded at week 13 post enrollment, and no further
data were generated for the subject.  If $R_1=0$,
$A_2 \in \mathcal{A}_{2,s}$ was generated as trinomial using the
current stage 2 randomization probabilities, and at week 25 post
enrollment, $Y_2$ was either generated as Bernoulli$\{p_2(a_1, a_2)\}$
for $A_1 = a_1, A_2=a_2$ if $Y_1=0$ or set equal to $Y_1$ if $Y_1=1$,
and $R_2$ was generated as Bernoulli$\{p_R(Y_2)\}$.  If $R_2=1$, $Y_2$
was recorded at week 26 post enrollment, and no further data were
generated for the subject.  If $R_2=0$, at week 38 post enrollment,
$Y_3$ was either generated as Bernoulli$\{p_3(a_1, a_2)\}$ for
$A_1 = a_1, A_2=a_2$ if $Y_2=0$ or set equal to $Y_2$ if $Y_2=1$.
Thus, $T_{end} = 143$ weeks.

\vspace*{0.1in}

\noindent
\textbf{D.2  Additional simulation results}

\noindent{\em Post-trial inference results $n_s=200$.}
Table~\ref{t:posttrial.supp} presents the same simulation results as in
Table~\ref{t:posttrial} of the main paper for Scenarios 2 and 4 with
$n_s=200$.  From the table, the results are all qualitatively similar
to those in the main paper.

\begin{table}
 \centering
 \caption{Post-trial results, $n_s = 200$, Scenarios 0, 2, and 4.
Entries are as in Table~\ref{t:posttrial} of the main paper.}
\label{t:posttrial.supp}
\begin{footnotesize}
 \begin{tabular}{l l c c c c c c c}    \Hline
Estimator & Measure & SR & TS(0.25) & TS(0.5) & TS(0.75) & TS(1) & TS($0.5t/T_{end}$)  & TS($t/T_{end}$) \\
   \hline
& & \multicolumn{7}{c}{Scenario 0, $\mu_s(a^{opt}_{1,s},a^{opt}_{2,s})$ = 0.521} \\*[0.03in]
$\mu_{s,Bayes}(a_1,a_2)$ & Est pCR Rate Opt Regime & 0.516 &0.511 &  0.504 &0.500 & 0.496 &  0.508& 0.500\\
                                      & Coverage                         & 0.949 & 0.954& 0.940 &0.945& 0.950 & 0.948& 0.953\\
                                      & Length                        & 0.287&0.289&0.292&0.298&0.305&0.290&0.296\\
                                     & Prop Est Correct              & 0.158 & 0.169& 0.173 & 0.174& 0.169 & 0.163& 0.161\\*[0.03in]
$\mu_{s,samp}(a_1,a_2)$ & Est pCR Rate Opt Regime & 0.518 & 0.511& 0.501 & 0.495& 0.487 & 0.507& 0.500\\
                                      & Coverage                         & 0.935 & 0.935& 0.917 & 0.918& 0.915 & 0.929& 0.925\\
                                      & Length                        & 0.310&0.312&0.316&0.324&0.331&0.313&0.321\\
                                     &  Prop Est Correct              & 0.159 & 0.167& 0.173 & 0.175& 0.168 & 0.165& 0.164\\
                                      & Rel Efficiency                   & 0.771 & 0.746& 0.682 & 0.635& 0.593 & 0.730& 0.662\\*[0.03in]
   $\mu_{s,wtsamp}(a_1,a_2)$ & Est pCR Rate Opt Regime & --& 0.516& 0.509 & 0.507& 0.501 & 0.514& 0.506\\
                                      & Coverage                         & -- & 0.938& 0.921 & 0.915& 0.904 & 0.930& 0.921\\
                                      & Length                        & --&0.314&0.322&0.334&0.331&0.313&0.321\\
                                     & Prop Est Correct              & -- &0.167 & 0.173 & 0.175& 0.168 & 0.165& 0.164\\
                                      &  Rel Efficiency               & -- & 0.761& 0.708 & 0.647& 0.590 & 0.741& 0.666\\*[0.05in]

          & & \multicolumn{7}{c}{Scenario 2, $\mu_s(a^{opt}_{1,s},a^{opt}_{2,s})$ = 0.712} \\*[0.03in]
$\mu_{s,Bayes}(a_1,a_2)$ & Est pCR Rate Opt Regime & 0.685 & 0.683 & 0.682 & 0.681& 0.677 & 0.682& 0.680\\
                                      & Coverage                         & 0.953 & 0.951& 0.938 & 0.930& 0.922 & 0.946& 0.930\\
                                     & Length                        & 0.266&0.243&0.229&0.220&0.218&0.236&0.223\\
                                    & Prop Est Correct              & 0.725 & 0.757& 0.769 & 0.789& 0.782 & 0.756& 0.776 \\*[0.03in]
$\mu_{s,samp}(a_1,a_2)$ & Est pCR Rate Opt Regime & 0.713 & 0.704& 0.700 & 0.698& 0.692 & 0.702& 0.697\\
                                      & Coverage                         & 0.923 & 0.943& 0.939 & 0.938& 0.930 &0.943 &0.936 \\
                                      & Length                        & 0.278&0.253&0.238&0.230&0.228&0.246&0.232\\
                                    &  Prop Est Correct              & 0.723 & 0.756& 0.769 & 0.789& 0.777 &0.759 & 0.773\\
                                      & Rel Efficiency                   & 0.898 & 1.038& 1.060 & 1.038& 1.003 & 1.056& 1.066\\*[0.03in]
   $\mu_{s,wtsamp}(a_1,a_2)$ & Est pCR Rate Opt Regime & --& 0.708& 0.706 & 0.704& 0.701 & 0.707& 0.703\\
                                      & Coverage                         & -- & 0.938& 0.939 &0.941 & 0.942 & 0.939& 0.941\\
                                      & Length                        & --&0.252&0.239&0.232&0.232&0.246&0.234\\
                                    & Prop Est Correct              & -- & 0.756& 0.769 & 0.789& 0.777 &0.759 & 0.773\\
                                      &  Rel Efficiency                  & -- &1.061 & 1.139 &1.147 & 1.153 & 1.105&1.162 \\*[0.05in]
   
   & & \multicolumn{7}{c}{Scenario 4, $\mu_s(a^{opt}_{1,s},a^{opt}_{2,s})$ = 0.712} \\*[0.03in]
$\mu_{s,Bayes}(a_1,a_2)$ & Est pCR Rate Opt Regime & 0.682 &0.682 &0.679&0.678&0.675&0.681&0.679\\
                                      & Coverage & 0.946 &0.944 &0.930&0.932&0.921&0.942&0.932\\
                                      & Length                        & 0.266&0.244&0.233&0.226&0.226&0.237&0.226\\
                                     & Prop Est Correct              & 0.695 &0.737&0.741&0.769&0.759&0.741&0.756\\*[0.03in]
$\mu_{s,samp}(a_1,a_2)$ & Est pCR Rate Opt Regime & 0.710 &0.704&0.698&0.695&0.691&0.701&0.696\\
                                      & Coverage & 0.928 &0.941&0.935&0.935&0.933&0.938&0.943\\
                                      & Length                        & 0.279&0.254&0.243&0.237&0.237&0.247&0.235\\
                                   &  Prop Est Correct & 0.704 &0.743&0.749&0.774&0.765&0.747&0.759\\
                                      & Rel Efficiency                   & 0.931&1.047&1.061&1.032&0.978&1.055&1.088\\*[0.03in]
   $\mu_{s,wtsamp}(a_1,a_2)$ & Est pCR Rate Opt Regime & --& 0.708&0.703&0.703&0.700&0.706&0.703\\
                                      & Coverage                         & -- & 0.940&0.938&0.937&0.936&0.936&0.946\\
                                      & Length                        & --&0.253&0.244&0.239&0.241&0.247&0.237\\
                                   & Prop Est Correct              & -- & 0.743&0.749&0.774&0.765&0.745&0.759\\
                                      &  Rel Efficiency                  & -- & 1.075&1.141&1.161&1.109&1.097&1.202\\*[0.05in]

   \hline
 \end{tabular}
 \end{footnotesize}
\end{table}

Table~\ref{t:effSR} shows Monte Carlo relative efficiency of the
indicated estimator under each RAR scheme to the
estimator under SR; because the natural frequentist estimator under SR
is $\hatmu_{s,samp}(a_1,a_2)$, the efficiency shown for
$\hatmu_{s,wtsamp}(a_1,a_2)$ under adaptive randomization is relative
to $\hatmu_{s,samp}(a_1,a_2)$ under SR.  Notably, the weighted
estimator under adaptive randomization is relatively more efficient in
most cases, with mild to moderate efficiency loss only under the most
aggressive schemes.  Efficiency of $\hatmu_{s,samp}(a_1,a_2)$ under
adaptive randomization relative to that under SR shows a similar but
less dramatic pattern.  The Bayesian estimator
$\hatmu_{s,Bayes}(a_1,a_2)$ shows only modest gains or losses under
less aggressive adaptation and greater losses than the other two
estimators under more aggressive adaptation, possibly a consequence of
its larger bias.  Overall, the results suggest that all estimators
perform well under less aggressive schemes, with
$\hatmu_{s,wtsamp}(a_1,a_2)$ gaining efficiency relative to estimation
under SR under these conditions, with no or only modest losses under
scenarios involving delayed effects.

\begin{table}
 \centering
 \caption{Monte Carlo relative efficiency of the estimator for the
   optimal regime for Scenarios 0 - 5 under adaptive randomization to
   the estimator under SR,  defined as Monte Carlo mean square error of the
      estimator under SR divided by that under the indicated adaptive
      randomization scheme. For $\mu_{s,wtsamp}(a_1,a_2)$, efficiency is
      relative to $\mu_{s,samp}(a_1,a_2)$ under SR.}
    \label{t:effSR}
    \begin{small}
 \begin{tabular}{l l c c c c c c}
   \Hline
Scenario & Estimator & TS(0.25) & TS(0.5) & TS(0.75) & TS(1) & TS($0.5t/T_{end}$)  & TS($t/T_{end}$) \\
   \hline
 0 & $\mu_{s,Bayes}(a_1,a_2)$ &  0.968 &0.827 &0.819 &0.770 & 0.944 &0.854\\
         & $\mu_{s,samp}(a_1,a_2)$  &  0.937 &0.731 &0.674 &0.592  &0.894 &0.734\\
   &  $\mu_{s,wtsamp}(a_1,a_2)$ & 0.956 &0.760 &0.687&0.589 & 0.907 &0.739\\*[0.03in]
   
1 & $\mu_{s,Bayes}(a_1,a_2)$ &  0.962&0.868 &0.800 &0.733  &0.930 &0.831\\
   & $\mu_{s,samp}(a_1,a_2)$  & 1.047 &0.953 &0.857 &0.765 & 1.017 &0.897\\
   &  $\mu_{s,wtsamp}(a_1,a_2)$ & 1.075 & 1.020 & 0.931& 0.844 & 1.051 &0.975\\*[0.03in]

2 & $\mu_{s,Bayes}(a_1,a_2)$ & 1.051 &0.997 &0.934 &0.820 & 1.034 &0.935\\
   & $\mu_{s,samp}(a_1,a_2)$  & 1.215 &1.178& 1.080& 0.917 & 1.216 &1.110\\
   &  $\mu_{s,wtsamp}(a_1,a_2)$ & 1.242& 1.266& 1.194& 1.054&  1.272& 1.211\\*[0.03in]

3 & $\mu_{s,Bayes}(a_1,a_2)$ & 0.962& 1.011 &0.802 &0.749 & 0.973& 0.872\\
   & $\mu_{s,samp}(a_1,a_2)$  & 1.083 &1.184 &0.890 &0.825 & 1.128 &1.000\\
   &  $\mu_{s,wtsamp}(a_1,a_2)$ & 1.115 &1.266 &0.997 &0.932&  1.184& 1.097\\*[0.03in]

4 & $\mu_{s,Bayes}(a_1,a_2)$ & 1.050 &0.941 &0.876 &0.776 & 1.035 &0.964\\
   & $\mu_{s,samp}(a_1,a_2)$  & 1.180& 1.072 &0.972 &0.815 & 1.173 &1.126\\
   &  $\mu_{s,wtsamp}(a_1,a_2)$ & 1.213 &1.153&1.093 &0.924 & 1.220  &1.245\\*[0.03in]

 5 & $\mu_{s,Bayes}(a_1,a_2)$ & 0.932 &0.824 &0.724 &0.634 & 0.879 &0.756\\
   & $\mu_{s,samp}(a_1,a_2)$  & 1.018& 0.903&0.778 &0.651 & 0.992&0.829\\
   &  $\mu_{s,wtsamp}(a_1,a_2)$ & 1.045 &0.964&0.855 &0.714 & 1.032 &0.903\\*[0.03in]
\hline
 \end{tabular}
 \end{small}
\end{table}

\vspace*{0.1in}

\noindent{\em ``Null'' scenario results, $n_s = 200$.}  Because it is
well known \citep{NorwoodRAR,zhang2021inference,ZhangMest2021} that
the post-trial inference after adaptive randomization using
frequentist techniques is particularly challenging when (in the
present context) there is no unique optimal embedded regime, we
consider a scenario reflecting the most extreme case where all regimes
achieve the same pCR rate, which we refer to as the null scenario,
Scenario 0.  Here, the probability of achieving pCR at the end of each
stage is the same for all treatment options; namely,
$p_1(a_1) = 0.30$, $a_1 = 0, 1$; $p_2(a_1, a_2) = 0.25,$
$(a_1, a_2) = (0,0), (0,1), (0, 2), (1,0), (1,1), (1,2)$; and
$p_3(a_1, a_2) = 0.15$,
$(a_1, a_2) = (0,0), (0,1), (0, 2), (1,0), (1,1), (1,2)$.  Under these
settings, the value of all six embedded regimes is equal to 0.521.
For definiteness, we arbitrarily take regime $\{ 0, 0\}$ to be the
optimal regime in calculation of the performance measures in the main
paper.  All other settings are as in the scenarios reported in the
main paper.

Table~\ref{t:nullin} shows the in-trial results for Scenario 0.  As
expected, adaptive randomization has no effect.  Post-trial inference
results for Scenario 0 are presented in Table~\ref{t:posttrial.supp}.
All estimators exhibit downward bias under all randomization schemes,
as for the scenarios reported in the main paper, with the smallest
bias occurring for $\hatmu_{s,samp}(a_1,a_2)$ under SR, where standard
asymptotic theory holds, and $\hatmu_{s,wtsamp}(a_1,a_2)$ under
adaptive randomization, as for the scenarios in the main paper.
Confidence intervals based on the frequentist estimators
$\hatmu_{s,samp}(a_1,a_2)$ and $\hatmu_{s,wtsamp}(a_1,a_2)$ exhibit
coverage below the nominal level in most cases, while those based on
$\hatmu_{s,Bayes}(a_1,a_2)$ achieve the nominal level; intervals based
on this estimator are also shorter $>$ 98\% of the time (not shown).
Interestingly, in contrast to the results in the more realistic
scenarios in the main paper, the frequentist estimators are relatively
much less efficient than $\hatmu_{s,Bayes}(a_1,a_2)$.  Finally,
Table~\ref{t:effSR} shows that under Scenario 0 all estimators can
exhibit notable efficiency loss under adaptive randomization relative
to SR, with $\hatmu_{s,Bayes}(a_1,a_2)$, although showing some loss,
faring considerably better than the other estimators.  Overall, these
results suggest that one base post-trial inference on this estimator
in settings where all regimes achieve similar values.

\begin{table}
 \centering
 \caption{In-trial results for Scenario 0, $n_s = 200$. Entries are as in
   Table~\ref{t:intrial} of the main paper.}
 \label{t:nullin}
 \begin{small}
 \begin{tabular}{l c c c c c c c c}   \Hline
Measure & Scenario & SR & TS(0.25) & TS(0.5) & TS(0.75) & TS(1) &
                                                                  TS($0.5t/T_{end}$)  & TS($t/T_{end}$) \\
   \hline
Overall pCR Rate   & 0 &  0.520 & 0.522 & 0.522 & 0.521 & 0.522 & 0.521 & 0.521\\

   Consist w/Opt & 0 & 0.243 & 0.243 & 0.244 & 0.244 & 0.242 & 0.244 & 0.243\\

Consist w/Worst   & 0 & 0.243 & 0.243 & 0.244 & 0.244 &  0.242 & 0.244 & 0.243\\

Rand Prob $a_1$ Opt & 0 & 0.500 & 0.500 & 0.500 & 0.505 & 0.497 & 0.501 &  0.499 \\

Rand Prob $a_2$ Opt & 0 & 0.333 & 0.333 & 0.333 & 0.335 &  0.334 & 0.333 & 0.335 \\*[0.03in]
 \hline
 \end{tabular}
 \end{small}
\end{table}

 % \begin{table}
%   \centering
%   \caption{Post-trial relative efficiency results, $n_s = 200$.  For
%     each estimator for the optimal regime in Scenario 0, Monte Carlo
%     relative efficiency of the estimator under adaptive randomization
%     to the estimator under SR (Monte Carlo mean square error of the
%     estimator under SR divided by that under the indicated adaptive
%     randomization scheme).  For $\mu_{s,wtsamp}(a_1,a_2)$, efficiency
%     is relative to $\mu_{s,samp}(a_1,a_2)$ under SR.}
%     \label{t:eff}
%  \begin{tabular}{l l c c c c c c}    \Hline
% Scenario & Estimator & TS(0.25) & TS(0.5) & TS(0.75) & TS(1) & TS($0.5t/T_{end}$)  & TS($t/T_{end}$) \\
%    \hline
%    0 & $\mu_{s,Bayes}(a_1,a_2)$ &  0.968 &0.827 &0.819 &0.770 & 0.944 &0.854\\
%          & $\mu_{s,samp}(a_1,a_2)$  &  0.937 &0.731 &0.674 &0.592  &0.894 &0.734\\
%    &  $\mu_{s,wtsamp}(a_1,a_2)$ & 0.956 &0.760 &0.687&0.589 & 0.907 &0.739\\*[0.03in]
%    \hline
%  \end{tabular}
% \end{table}

\vspace*{0.1in}

\noindent{\em Scenario 3 results, $n_s=120$ and $n_s=1000$.}
Table~\ref{t:s3intrial} presents in-trial results for Scenario 3 under
the smaller sample size $n_s= 120$, which would be considered
``small'' in practice for a SMART involving six embedded regimes, and
$n_s=1000$, chosen as a sample size that, while possibly larger than
resources might allow in some settings, represents a situation where
large sample theory would be likely provide good approximations to
finite-sample performance.  Not surprisingly, for $n_s=120$, gains in
overall pCR rate under RAR are negligible, as are increases in the
proportions of subjects consistent with the optimal regime.
Randomization probabilities under RAR do increase with more aggressive
adaptation, particularly for stage 1, but apparently not sufficiently
to produce gains in the previous measures.  In contrast, for
$n_s=1000$, gains of up to 5\% in the overall pCR rate are obtained
under RAR, with roughly 50\% of subjects consistent with the optimal
regime under more aggressive versions.  This feature is undoubtedly a
consequence of the very high randomization probabilities under RAR for
the stage 1 and 2 treatments associated with the optimal regime.
Evidently, with the rich information from this large sample, the RAR
methods are able to identify definitively the optimal regime and assign
subjects to it.

Table~\ref{t:s3posttrial} shows post-trial results.  With $n_s=120$,
performance is degraded for the Bayesian estimator, which is very
downward biased, whereas the bias is much more modest under
$n_s=1000$, demonstrating that this bias is in part a sample size
issue.  Likewise, the downward bias of the weighted estimator is mile
for $n_2=120$ and almost eliminated entirely with $n_s=1000$, again
demonstrating that this bias is tied to sample size.  Coverage
probabilities mostly fall short of the nominal 0.95 level for
$n_s=120$, reflecting the considerable uncertainty; for $n_s=1000$,
coverage achieves the nominal level.  The length of intervals based on
the Bayesian estimator is shorter than the others 85-90\% of the time
for $n_s=120$ and about 80\% of the time for $n_2=1000$ (not shown).  The
estimators $\hatmu_{s,Bayes}(a_1,a_2)$ and
$\hatmu_{s,wtsamp}(a_1,a_2)$ estimators are mostly equivalent in terms
of efficiency under RAR with both sample sizes.

Table~\ref{t:effSRns} shows that efficiency relative to SR is either
the same or worse under $n_s=120$ for all RAR methods.  With the large
sample size $n_s=1000$, considerable gains in efficiency over SR under
RAR are possible with all estimators.

\begin{table}
 \centering
 \caption{In-trial results for Scenario 3, $n_s = 120$ and $n_2=1000$. Entries are as in
   Table~\ref{t:intrial} of the main paper.}
 \label{t:s3intrial}
 \begin{small}
 \begin{tabular}{l c c c c c c c c}   \Hline
Measure & Scenario & SR & TS(0.25) & TS(0.5) & TS(0.75) & TS(1) &
                                                                  TS($0.5t/T_{end}$)  & TS($t/T_{end}$) \\
   \hline
& & \multicolumn{7}{c}{$n_s=120$}\\*[0.03in]   
Overall pCR Rate   & 3 &  0.590 & 0.597 & 0.602 & 0.607 & 0.610 & 0.600 & 0.606\\
   Consist w/Opt & 3 & 0.244 & 0.272 & 0.301 & 0.321 & 0.338 & 0.284 & 0.316\\
Consist w/Worst   & 3 & 0.257 & 0.241 & 0.203 & 0.212 &  0.203 & 0.233 & 0.217\\
Rand Prob $a_1$ Opt & 3 & 0.500 & 0.585 & 0.660 & 0.706 & 0.740 & 0.656 &  0.737 \\
Rand Prob $a_2$ Opt & 3 & 0.333 & 0.407 & 0.470 & 0.504 &  0.537 & 0.465 & 0.546 \\*[0.03in]

& & \multicolumn{7}{c}{$n_s=1000$}\\*[0.03in]   
Overall pCR Rate   & 3 &  0.590 & 0.620 & 0.636 &0.643 & 0.647 & 0.625 & 0.638\\
   Consist w/Opt & 3 & 0.243 & 0.385 & 0.475 & 0.523 & 0.544 & 0.416 & 0.496\\
Consist w/Worst   & 3 & 0.252 & 0.178 & 0.142 & 0.126 &  0.119 & 0.168 & 0.138\\
Rand Prob $a_1$ Opt & 3 & 0.500 & 0.836 & 0.920 & 0.940 & 0.946 & 0.922 &  0.947 \\
Rand Prob $a_2$ Opt & 3 & 0.333 & 0.608 & 0.733 & 0.801 &  0.826 &  0.723 & 0.824 \\*[0.03in]
 \hline
 \end{tabular}
 \end{small}
\end{table}

\begin{table}
 \centering
 \caption{Post-trial results, Scenario 3, $n_s = 120$ and
   $n_2=1000$. $\mu_s(a^{opt}_{1,s},a^{opt}_{2,s})$ = 0.712 for regime
   $\{0, 0\}$.
Entries are as in Table~\ref{t:posttrial} of the main paper.}
\label{t:s3posttrial}
\begin{footnotesize}
 \begin{tabular}{l l c c c c c c c}    \Hline
Estimator & Measure & SR & TS(0.25) & TS(0.5) & TS(0.75) & TS(1) & TS($0.5t/T_{end}$)  & TS($t/T_{end}$) \\
   \hline
& & \multicolumn{7}{c}{$n_s=120$} \\*[0.03in]
$\mu_{s,Bayes}(a_1,a_2)$ & Est pCR Rate Opt Regime & 0.670 &0.667 &  0.665 &0.658 & 0.656 &  0.665& 0.662\\
                                      & Coverage                         & 0.952 & 0.940& 0.940 &0.924& 0.924 & 0.938& 0.929\\
                                      & Length                   & 0.331&0.317&0.306&0.304&0.302&0.312&0.303\\
                                   & Prop Est Correct              & 0.536 & 0.556& 0.584 & 0.570& 0.579 & 0.561& 0.594\\*[0.03in]
$\mu_{s,samp}(a_1,a_2)$ & Est pCR Rate Opt Regime & 0.712 & 0.702& 0.696 & 0.684& 0.680 & 0.699& 0.690\\
                                      & Coverage                         & 0.911 & 0.927& 0.931 & 0.925& 0.927 & 0.930& 0.928\\
                                      & Length                   & 0.355&0.342&0.332&0.333&0.329&0.337&0.328\\
                                     &  Prop Est Correct              & 0.546 & 0.571& 0.595 & 0.578& 0.576 & 0.569& 0.599\\
                                      & Rel Efficiency                   & 0.843 & 0.932& 0.967 & 1.014& 0.860 & 0.961& 0.918\\*[0.03in]
   $\mu_{s,wtsamp}(a_1,a_2)$ & Est pCR Rate Opt Regime & --& 0.706& 0.705 & 0.695& 0.693 & 0.705& 0.700\\
                                      & Coverage                         & -- & 0.924& 0.930 & 0.925& 0.922 & 0.928& 0.922\\
                                      & Length                   & --&0.339&0.330&0.332&0.330&0.335&0.328\\
                                      & Prop Est Correct              & -- &0.571 & 0.595 & 0.578& 0.576 & 0.569& 0.599\\
                                      &  Rel Efficiency               & -- & 0.957& 1.037 & 1.014& 0.953 & 1.006& 0.990\\*[0.05in]

          & & \multicolumn{7}{c}{$n_s=1000$} \\*[0.03in]
$\mu_{s,Bayes}(a_1,a_2)$ & Est pCR Rate Opt Regime & 0.706 & 0.705 & 0.706 & 0.706& 0.706 & 0.705& 0.706\\
                                      & Coverage                         & 0.948 & 0.942& 0.948 & 0.944& 0.938 & 0.946& 0.937\\
                                      & Length                   & 0.125&0.096&0.086&0.082&0.081&0.092&0.084\\
                                      & Prop Est Correct              & 0.885 & 0.926& 0.945 & 0.951& 0.942 & 0.935& 0.944 \\*[0.03in]
$\mu_{s,samp}(a_1,a_2)$ & Est pCR Rate Opt Regime & 0.712 & 0.709& 0.709 & 0.709& 0.709 & 0.709& 0.709\\
                                      & Coverage                         & 0.947 & 0.947& 0.948 & 0.948& 0.940 &0.950 &0.943 \\
                                      & Length                   & 0.126&0.096&0.086&0.082&0.081&0.092&0.084\\
                                   &  Prop Est Correct              & 0.884 & 0.927& 0.946 & 0.953& 0.942 &0.936 & 0.944\\
                                      & Rel Efficiency                   & 0.989 & 1.067& 1.060 & 1.069& 1.065 & 1.066& 1.069\\*[0.03in]
   $\mu_{s,wtsamp}(a_1,a_2)$ & Est pCR Rate Opt Regime & --& 0.710& 0.710 & 0.710& 0.710 & 0.710& 0.710\\
                                      & Coverage                         & -- & 0.948& 0.951 &0.951 & 0.952 & 0.954& 0.947\\
                                      & Length                   & --&0.098&0.088&0.085&0.084&0.094&0.087\\
                                   & Prop Est Correct              & -- & 0.927& 0.946 & 0.953& 0.942 &0.936 & 0.944\\
                                      &  Rel Efficiency                  & -- &1.065 & 1.066 &1.089 & 1.121 & 1.068&1.086 \\*[0.05in]
   
    \hline
 \end{tabular}
 \end{footnotesize}
\end{table}

\begin{table}
 \centering
 \caption{Monte Carlo relative efficiency of the estimator for the
   optimal regime under adaptive randomization to
   the estimator under SR, defined as in Table~\ref{t:effSR}, Scenario
 3, $n_s=120$ and $n_s=1000$}
    \label{t:effSRns}
    \begin{small}
 \begin{tabular}{l l c c c c c c}
   \Hline
$n_s$ & Estimator & TS(0.25) & TS(0.5) & TS(0.75) & TS(1) & TS($0.5t/T_{end}$)  & TS($t/T_{end}$) \\
   \hline
 120 & $\mu_{s,Bayes}(a_1,a_2)$ &  0.929 &0.889 &0.757 &0.699 & 0.905 &0.796\\
         & $\mu_{s,samp}(a_1,a_2)$  &  1.028&1.019 &0.833 &0.713  &1.032 &0.867\\
   &  $\mu_{s,wtsamp}(a_1,a_2)$ & 1.055 &1.094 &0.911&0.790 & 1.080 &0.936\\*[0.035in]
   
1000 & $\mu_{s,Bayes}(a_1,a_2)$ &  1.461&1.771 &1.734 &1.558  &1.562 &1.671\\
   & $\mu_{s,samp}(a_1,a_2)$  & 1.577 &1.900 &1.875 &1.678 & 1.684 &1.806\\
   &  $\mu_{s,wtsamp}(a_1,a_2)$ & 1.574 & 1.909 & 1.910& 1.766 & 1.688 &1.834\\*[0.03in]
\hline
 \end{tabular}
 \end{small}
\end{table}

\end{document}